\documentclass[11pt]{article}

\usepackage{geometry} 
\usepackage{import}
\usepackage[parfill]{parskip}    
\usepackage{graphicx}
\usepackage{amssymb}
\usepackage{amsmath}
\DeclareMathOperator*{\argmax}{arg\,max}
\usepackage{ulem}
\usepackage{rotating}           	
\usepackage{eurosym}
\usepackage{multirow}				
\usepackage{epstopdf}
\usepackage{authblk}
\usepackage{xcolor}
\usepackage[noend]{algpseudocode}
\usepackage{algorithm}
\usepackage{titlesec}
\usepackage{longtable}
\usepackage{booktabs}
\usepackage{enumitem}
\usepackage{caption}
\usepackage{subcaption}
\usepackage{hyperref}
\usepackage{mathrsfs}
\usepackage{array}
\usepackage{arydshln} 
\usepackage{lscape} 
\usepackage{xcolor}
\usepackage{tikz}
\usepackage{cleveref}
\usetikzlibrary{fadings}
\usetikzlibrary{
    decorations.pathreplacing, 
    angles, 
    quotes, 
    matrix, 
    fit, 
    positioning, 
    arrows.meta,
    decorations.pathmorphing, 
    decorations.shapes,
    shadings,
    fadings,
}
\usepackage[utf8]{inputenc}
\usepackage{soul}
\usepackage[flushleft]{threeparttable}
\newcolumntype{T}{>{\footnotesize}l} 
\newcolumntype{F}{>{\footnotesize}c} 
\newcolumntype{R}{>{\footnotesize}r} 
\usepackage{natbib}
\definecolor{darkgreen}{rgb}{0.16,0.67,0.17}
\usepackage{amsthm}
\newtheorem{property}{Property}
\newtheorem{observation}{Observation}
\allowdisplaybreaks

\usepackage{fancyhdr}
\titleformat{\paragraph}
{\normalfont\normalsize\bfseries}{\theparagraph}{1em}{}
\titlespacing*{\paragraph}
{0pt}{3.25ex plus 1ex minus .2ex}{1.5ex plus .2ex}

\graphicspath{{Supplementary/}}
\hypersetup{
	colorlinks=true,
	linkcolor=blue,
	citecolor=blue,
	filecolor=magenta,      
	urlcolor=blue,}
\makeatletter
\def\BState{\State\hskip-\ALG@thistlm}
\makeatother

\algrenewcommand\algorithmicforall{\textbf{foreach}}
\algrenewcommand\algorithmicindent{.8em}
\definecolor{darkgreen}{rgb}{0.16,0.67,0.17}
\begin{document}
\title{A Bilevel Approach to Integrated Surgeon Scheduling and Surgery Planning solved via Branch-and-Price}

\author[1,2] {\normalsize Broos Maenhout}
\author[3] {P\v remysl \v S\r ucha}
\author[3] {Viktorie Valdmanov\' a}
\author[3] {Ond\v rej Tkadlec}
\author[3] {Jana Thao Rozlivkov\' a\footnotesize}
\affil[1]{Faculty of Economics and Business Administration, Ghent University, Tweekerkenstraat 2, 9000 Gent (Belgium), Broos.Maenhout@Ugent.be}
\affil[2]{\footnotesize FlandersMake@UGent – corelab CVAMO, Ghent, Belgium}
\affil[3] {Czech Institute of Informatics, Robotics and Cybernetics, Czech Technical University in Prague, Jugoslávských partyzánů 1580/3, 160 00 Prague, (Czech Republic), suchap@cvut.cz, valdmvik@fel.cvut.cz, ondra.tkadlec.10@seznam.cz, janathaorozlivkova@gmail.com}

\date{--------------------------------------------------------}
\setlength{\parindent}{0 pt}	
\maketitle

\begin{abstract}
In this paper, we study a multi-agent scheduling problem for organising the operations within the operating room department. The head of the surgeon group and individual surgeons are together responsible for the surgeon schedule and surgical case planning. The surgeon head allocates time blocks to individual surgeons, whereas individual surgeons determine the planning of surgical cases independently, which might degrade the schedule quality envisaged by the surgeon head. The bilevel optimisation under study seeks an optimal Nash equilibrium solution -- a surgeon schedule and surgical case plan that optimise the objectives of the surgeon head, while ensuring that no individual surgeon can improve their own objective within the allocated time blocks. We propose a dedicated branch-and-price that adds lazy constraints to the formulation of surgeon-specific pricing problems to ensure an optimal bilevel feasible solution is retrieved. In this way, the surgeon head respects the objective requirements of the individual surgeons and the solution space can be searched efficiently. In the computational experiments, we validate the performance of the proposed algorithm and its dedicated components and provide insights into the benefits of attaining an equilibrium solution under different scenarios by calculating the price of stability and the price of decentralisation.

\end{abstract}

\section{\label{section_introduction}Introduction}
The Operating Room (OR) department is of major importance in the activities and profitability of a hospital. \textcolor{black}{Its more effective utilisation is likely to decrease costs for surgical care delivery, improve patient access and wait times, and increase the volume of cases treated} \citep{Roshanaei2017}. However, in real-life environments, utilisation is hampered due to inefficient scheduling systems, partly caused by conflicting interests of the different stakeholders. The organisation of resources in the OR department is further complicated by the nature of this decision-making process, which is hierarchical and typically undertaken in three phases \citep{Batun2011,Fugener2014}. The \textit{first phase}, at the strategic level, considers the case-mix planning, involving the distribution of OR time between surgical disciplines. The \textit{second phase}, at the tactical level, encompasses the composition of the master surgery schedule (MSS), which assigns blocks of consecutive time slots in a specific OR on a particular day to a group of surgeons, typically related to a particular discipline \citep{VanOostrum2012}. The OR blocks that are assigned in the MSS will be reserved for the relevant surgery types associated with the allotted surgeon group. This scheduling strategy is also referred as block scheduling. The \textit{third phase} concerns the operational planning and scheduling of patients and resources and involves (i) the composition of a short-term surgeon schedule that assigns adjacent OR time blocks to individual surgeons based on their patient waiting lists and, if applicable, the MSS; (ii) advance patient planning that determines the surgery date of patients; and (iii) surgery scheduling that sequences the patients in the assigned ORs. The classical approach to mimic the OR planning and scheduling process is to employ a centralised hierarchical approach solving these multiple phases having different objectives separately, for which the decisions of higher hierarchical stages constrain the decisions of lower stages, or to formulate a multi-level integrated decision problem with multiple weighted objectives \citep{VanRiet2015}. However, these approaches do not reflect real-life practice as the planning and sequencing of surgeries in the operational phase are performed in a decentralised manner, by the involved surgeons according to their individual judgements \citep{Johnston2019}. The latter may compromise the outcomes and objectives set by the OR manager, necessitating a careful optimisation approach. 

In this paper, we study a multi-agent surgeon scheduling problem, at the operational level, which is interrelated with the advance planning of surgeries. The problem under study has been observed in real life and assumes the block booking scheduling framework, which assigns eligible time blocks of varying lengths to surgeons. The surgeries of patients are typically planned one or two weeks before their execution, considering a smaller horizon of, for example, one week. The creation of a short-term surgeon timetable is required to deliver a feasible surgeon schedule and organise the planning of surgeries efficiently. Individual surgeons communicate their requests for OR time for the upcoming (short-term) period to the head of the surgeons' group, based on the set of patients on their waiting lists. The surgeon head distributes the available OR time, determined in higher-level strategic and tactical decision levels, to the individual surgeons. The surgeons, on the other hand, are responsible for deciding on the allocation of patients to days in the planning horizon. The resulting surgeon schedule and surgery planning are typically the result of a negotiation between the surgeon head and the (multiple) individual surgeons. In the bilevel problem under study, the \textit{surgeon head} is identified as the \textit{leader}, standing at the top of the team hierarchy. His objective is to maximise OR utilisation and the case volume treated, possibly weighting individual patients based on considerations such as waiting time, duration, and revenue. The \textit{individual surgeons} are the \textit{followers}, who are at a lower layer of the hierarchy, addressing the planning of patients separately in an independent manner. Their objective is to maximise the patient priority of individual cases treated, taking their perspectives into account, which can be different from the priorities set by the surgeon head. The interaction between both agents is hierarchical in the sense that the realised outcome of any distribution of OR time decided by the head of the surgeon group is imposed to the lower-level individual surgeons, planning surgeries within the allotted time. When the higher-level decision-maker optimises its objectives independently, the envisaged outcome may be affected by the responses of the lower-level decision-makers. 
To tackle this problem, we propose a bilevel framework that encapsulates the interplay between the surgeon head and the individual surgeons to realise a consensus on the resulting surgeon schedule and patient planning. This approach, which integrates both problems in a specific manner, allows to find a stable solution representing a Nash equilibrium. The case planning decisions of individual surgeons are modelled as a nested inner optimisation problem that is included in the form of dynamically generated constraints in the outer optimisation problem related to the decision of the surgeon head.  

The problem under study is complex, and due to the size of real-life instances and large number of symmetric solutions, the compact mixed-integer programming (MIP) formulation based on the original assignment variables is inefficient. To address this issue, we apply the Dantzig-Wolfe decomposition, resulting in a master problem formulation that utilises pattern variables to represent feasible schedules encompassing the OR time schedule and patient planning of individual surgeons. While column-based formulations are typically effective for solving large-scale instances, developing a viable solution methodology remains challenging. To tackle this, we propose a dedicated branch-and-price algorithm that efficiently finds an optimal integer solution within a short time frame. A column generation algorithm solves the linear programming relaxation (LPR) of the leader problem, which satisfies all upper and lower-level constraints but relaxes the integer domain conditions. If the resulting optimal LPR solution is fractional, a branching scheme is applied to derive an integer solution. To ensure a bilevel feasible solution is retrieved, additional so-called lazy constraints are added via feasibility callbacks to the formulation of surgeon-specific pricing problems, representing feasible schedules of individual surgeons. In this way, the leader has to respect the objective requirements of the followers and the solution space of the leader problem is restricted. The performance of the algorithm has been improved via different dedicated techniques, including the formulation and remembering of strengthened lazy constraints, the generation of multiple patterns per iteration, and the initialisation via a constructive heuristic, which produces bilevel feasible solutions. In the computational experiments, we demonstrate the impact of different proposed speed-up mechanisms and compare different lazy-constraint formulations. In addition, we validate the benefits of attaining an equilibrium solution under different scenarios by calculating the degradation of the objective due to selfish behaviour of the individual surgeons (\textit{price of decentralisation}) and degradation of the objective due to the required equilibrium (\textit{price of stability}).

The remainder of this paper is structured as follows. Section~\ref{section_literature} discusses the relevant literature. Section~\ref{section_problem} explains the multi-agent surgeon scheduling problem under study and presents a mathematical formulation of the resulting bilevel program. Section~\ref{section_methodology} gives insight into the formulation of lazy constraints, required to retrieve bilevel feasible solutions. Section~\ref{section_BaP} discusses the proposed branch-and-price procedure for finding optimal solutions. In Section~\ref{section_results}, we validate the performance of the proposed solution methodology and give insight into the price of stability and the price of decentralisation for different parameter settings and scenarios. Conclusions and directions for future research are provided in Section~\ref{section_conclusion}.

\section{\label{section_literature}Literature review}
Over the past years, a wide variety of OR scheduling problems has been studied in different guises and forms. For an exhaustive overview, we refer to recent overview papers of \cite{Zhu2019}, \cite{Savva2019}, and \cite{Rahimi2021}, amongst others. In this literature review, we focus on the integrated allocation of individual surgeons to blocks and advance patient planning (Section \ref{subsection_OR_scheduling}). In addition, we review the applications of bilevel optimisation -- primarily focusing on healthcare -- to derive equilibrium solutions taking the objectives of different stakeholders into account (Section \ref{subsection_bilevel}). Lastly, we highlight the contributions of our research (Section \ref{subsection_literature_contributions}).


\subsection{\label{subsection_OR_scheduling}Integrated surgeon scheduling and surgical case advance planning}


Operational decisions concern short-term allocation of individual surgeons to blocks and planning and scheduling of elective patients. The majority of studies considers these problems separately. However, different papers studying the planning of patients acknowledge the importance of the resource schedule and consider constraints related to human resource availability or surgeon preferences to design the patient planning \citep[e.g.,][]{Roland2010, Meskens2013}. Only few studies address the integrated planning and scheduling decisions involved in solving a weekly planning problem. This includes the assignment of OR blocks to surgeons and the planning of patients for surgery in each block. Online Appendix A provides a summary table of the studies most relevant to our research, based on the characteristics of the tackled problems. Analysis focuses on the considered types of decisions and the decision granularity. A surgeon scheduling problem may assign surgeons to days, rooms, blocks, and time slots, whereas the patient allocation problem may concern the advance planning and/or sequencing of patients. For example, \cite{Vanhuele2014} consider the integrated surgeon scheduling, advance planning and surgical case sequencing in an open booking framework based on surgeon preferences. The granularity for allocating OR time to individual surgeons has typically been predefined in advance and varies across studies. It can involve fixed-length blocks spanning an entire OR day \citep{Agnetis2014}, half a day \citep{Penn2017}, or a work shift \citep{Dios2015a}. Other authors let the OR time allocated to surgeons depend upon the scheduling of individual surgeries by defining their start times \citep[e.g.,][]{ValiSiar2018,Younespour2019}. The latter is typically the case when the advance planning problem is integrated with the patient sequencing problem \citep[e.g.,][]{Vanhuele2014,Akbarzadeh2019}. In this study, we integrate the advance planning of patients and the short-term (e.g., weekly) scheduling of surgeons, distributing the available time allotted to a discipline in the MSS among the involved surgeons. This is achieved by splitting tactical MSS blocks into operational surgeon blocks. Unlike previous studies that commonly consider the allocation of blocks having fixed lengths, this study considers blocks having variable lengths and start times, which is guided by the surgeons' caseload and characteristics of involved patients. This approach is informed by empirical observations and aligns with the findings of \cite{Akbarzadeh2025}.\\
In previous studies, different patient- and resource-related performance measures have been investigated reflecting the preferences and interests of involved stakeholders such as the OR manager, hospital management, individual surgeons, patients, or postoperative services \citep{Samudra2016,Gur2018}. An essential feature in many studies is the definition of patient priorities, which steer both the selection of patients for surgery within the considered horizon and the distribution of OR time between surgeons. Patient priorities may be set in accordance with financial objectives or medical considerations, amongst other, and are determined based on the preferences of the OR manager and/or involved surgeons \citep[e.g.,][]{Lamiri2008,Guda2016, Lotfi2022}. Apart from patient priorities, surgical case plans have been evaluated based upon the case volume planned \citep[][]{Spratt2016, Guido2017}, patients’ waiting times \citep{Testi2009, Kamran2018a, Hamid2019}, preferences of surgeons \citep{Roland2010, DiMartinelly2014}, and utilisation of OR resources such as rooms and surgeons \citep[][]{Penn2017,Hamid2019}. As a result, many studies consider a multi-objective problem accounting for the interests of multiple stakeholders. These objectives are handled either via the definition of a single weighted objective function or via the definition of multiple single- or multiple-objective optimisation models that are sequentially solved, as indicated in the summary table in Online Appendix A. These traditional approaches typically assume a single decision-maker who adopts a particular (unified) perspective, possibly combining the interests of multiple stakeholders. However, such models are inherently one-sided, as they overlook the complex interaction between the multiple autonomous actors, such as between the head surgeon and individual surgeons when composing the weekly surgeon schedule. While rational from their standpoint, the decisions of these individual surgeons may result in suboptimal outcomes and inefficiencies. In this study, however, we acknowledge the interplay of multiple autonomous actors, which better reflects the real-life decision-making context. We apply bilevel optimisation modelling the hierarchical relationship between a central authority and individual surgeons, facilitating coordination of decisions and identification of a stable solution between all parties.

Previous studies in the literature have demonstrated it is challenging to find high-quality solutions when solving problems related to surgeon scheduling and advance patient planning in real-life settings, because of the large number of decision variables and specific combination of objectives and constraints \citep{Marynissen2019}. Both exact and heuristic methods have been proposed, typically relying on a decomposed problem formulation and mathematical programming \citep{Samudra2016}. Column generation has been regularly applied to efficiently find high-quality solutions for patient planning problems \citep[e.g.,][]{Lamiri2008, Fei2008,Doulabi2016}. These studies rely on either heuristic principles or an exact approach to transform an optimal linear programming solution into an integer solution. \cite{Lamiri2008} and \cite{Akbarzadeh2019} propose a multi-phased heuristic solution methodology. They first apply column generation to solve the relaxed linear programming model and use the yielded optimal solution to build a feasible integer solution. \cite{Fei2009} and \cite{Akbarzadeh2020} also apply column generation to solve the relaxed problem and subsequently present a diving heuristic to find a feasible integer solution of high quality. \cite{Doulabi2016} propose combining column generation and branching, known as branch-and-price, with different speed-up techniques to accelerate the procedure for finding a high-quality solution. \cite{Koutecka2025} study the integrated advance planning and sequencing of surgeries and propose a branch-and-price approach that exploits machine learning for the ranking of pricing problems to enhance its efficiency. Also other methodologies frequently rely on some type of decomposition, proposing to solve individual, smaller problems with a singular objective and relevant constraints in a sequential manner \citep[][]{Vanhuele2014, Mazloumian2022}. For example, \cite{Agnetis2014} propose a decomposition approach focusing first on assigning the different surgical disciplines to sessions and then allocating surgeries to each session to determine the surgeon schedule on a weekly basis. Similarly, \cite{Aringhieri2015} propose a two-level tabu search meta-heuristic, iterating between the two problems in such a way as to address each one individually, while trying to globally improve the solution. Similar to previous studies, we apply a decomposition approach for solving the integrated problem. More precisely, we propose an exact branch-and-price method that relies on Dantzig-Wolfe decomposition and different problem properties to find optimal solutions in a reasonable time span.

\subsection{\label{subsection_bilevel}Bilevel optimisation}

Bilevel programming has been proven to be relevant for modelling real-world problems that can be interpreted as a hierarchical game of two decision makers and transcends single-level optimisation \citep{Colson2007}. Unlike classical single-level optimisation problems, the agents have their own objectives that need to be respected to come to an agreement. Such problems need to be modelled using game-theoretic approaches, for which it is essential to derive \textit{stable solutions}, i.e., a solution where the agents do not have an incentive to deviate from it. A stable solution guarantees relative satisfaction of (a) \textit{follower} agent(s) while maximising a global objective defined by a central authority or \textit{leader} agent. The leader decides first and conveys his/her decision to the follower agent(s), who optimise own objectives given the decision imposed by the leader, possibly impacting the leader’s reward. However, the leader has knowledge about the followers’ decision problem and, therefore can \textcolor{black}{anticipate the followers’ behaviour} in his/her decision-making accordingly \citep{Kalashnikov2016}. Hence, a bilevel program is an optimisation problem where the feasible set of decisions taken by the \textit{leader} is partly determined through a solution set mapping of second-level parametric optimisation problems modelling the behaviour of the follower(s) \citep{Dempe2002}. The two levels of the model are hierarchically organised and the upper leader's level communicates the variables it controls to the lower followers' level. Note that, due to the lack of computationally affordable methods, most studies in the literature assume only a single follower, which is addressed much more than the case with multiple followers \citep{Basilico2020}.\\ 
Bilevel optimisation has been shown useful to model applications involving electricity markets, transportation problems, and supply chains. However, its application to healthcare-related problems is limited. \cite{Chen2021} consider the home healthcare problem and formulate a leader's model to assign customers to nurses with as objective the minimisation of the total operating cost, whereas the follower's model is conceived as a routing problem maximising the degree of patient satisfaction with the visit time. In the same context of the home healthcare problem, \cite{Fathollahi2022} study a location-allocation-routing decision problem and consider pharmacies or nurses as the leader and all patients together as the follower entity. \cite{Zhang2010} consider a similar location-allocation problem to improve the accessibility of preventive healthcare facilities to potential clients and maximise participation. The upper level considers a capacitated facility location problem whereas the lower level determines the allocation of clients to facilities. \cite{Cao2023} and \cite{Valizadeh2023} study location-transport integrated optimisation problems. \cite{Cao2023} propose a bilevel program to improve the strategic decision of temporary disposal centres and the operational transport decisions to remove COVID-19 medical waste. \cite{Valizadeh2023} investigate multi-period COVID-19 vaccine location and distribution strategies to decelerate the coronavirus transmission.  \cite{Zhou2016} propose a bilevel optimisation problem to solve an outpatient appointment scheduling problem based on revenue management. The hospital, the leader in the hierarchy, decides the mix of physician resources hired to maximise the total profit. The outpatient department is the follower, which determines the appointment schedule to maximise its own profit. None of the previous studies applying bilevel optimisation have explored its use in planning and scheduling OR processes. In this study, we model the negotiation between the surgeon head and individual surgeons related to a specific discipline to coordinate the distribution of OR time effectively.

Techniques for solving bilevel optimisation problems were recently summarised in an extensive survey by \cite{Kleinert2021}, who indicate a wide variety of exact, heuristic, and meta-heuristic solution approaches. The preferred approach depends strongly on the domain of the variables (\textit{continuous} versus \textit{integer}) in the leader and follower problem. A common methodology is to reformulate a bilevel linear program with continuous lower-level variables into a single-level equivalent using linear programming duality or the Karush-Kuhn-Tucker conditions \citep{Kalashnikov2016, Kleinert2021}. Imposing integrality conditions on the variables produces a parametric mixed-integer program, possibly leading to a disconnected search space. The algorithms developed for solving discrete bilevel problems are usually dedicated and make use of specific features of the problem under study, relying on (specialised) branching schemes, addition of cutting planes, or approximation methods. \cite{Moore1990} and \cite{Bard1992} propose a generic \textit{branch-and-bound method} for discrete bilevel optimisation, pointing out that some of the standard branch-and-bound fathoming rules for MIP are not valid in the bilevel context. This algorithm has been extended in later studies, which propose more efficient algorithms thriving on multi-way branching \citep[][]{Xu2014} or branch-and-cut \citep[e.g.,][]{DeNegre2009, Fischetti2017}. Another frequently used method for solving binary bilevel optimisation problems is \textit{parametric programming}  \citep{Faisca2007,Koppe2010}. These methods enumerate first a large number of lower-level solutions, which are subsequently plugged into the upper-level problem, defining a single-level optimisation problem. Another relevant method to solve discrete bilevel programming problems via generation of cutting planes are \textit{Benders decomposition} that derives valid Benders-like cuts by fixing values at the master level \citep[e.g.,][]{Saharidis2009}. In the latter perspective, \cite{Milicka2022} propose to combine mathematical programming and lazy constraint generation. Lazy constraints are able to remove integer-feasible solutions, in contrast to cutting planes that cut off non-integer solutions from the LPR of the original problem. The study of \cite{sucha2021}, addressing multi-agent project scheduling, shows that this approach can be significantly faster on larger instances compared to an approach based on linear programming duality. Similar to these latter approaches, this study leverages the generation of lazy constraints. Notably, we are the first to propose a branch-and-price procedure for finding optimal, bilevel feasible solutions. Our approach offers two major advantages. First, the associated Dantzig-Wolfe decomposition ensures the generation of only bilevel feasible columns, representing viable schedules for the follower agents. Accordingly, the corresponding linear relaxation provides a stronger bound compared to using a traditional branch-and-bound approach that relies on the original assignment variables. Second, as each pattern is inherently bilevel feasible, there is no need to branch on the follower variables once the leader variables are determined, further improving the efficiency of the optimisation search. Consequently, the proposed approach results in a significantly smaller branching tree.

\subsection{\label{subsection_literature_contributions}Contributions of this study}

The contribution of this paper is threefold. First, we introduce a new optimisation model representing the hierarchical decision-making process related to the distribution of OR time, scheduling of surgeons, and planning of surgical cases, modelling the interaction between distinct stakeholders. The model optimises the objectives of the surgeon head and ensures a stable solution that cannot be affected by the behaviour of individual surgeons as they are not able to improve their own solution given the allocated OR time. Second, we propose an exact branch-and-price approach to exploit the associated decomposed problem formulation for finding bilevel feasible solutions efficiently, i.e., the pricing procedure adds new cutting planes to the subproblem formulation to derive an optimal solution. In this way, only relevant surgical case plans for individual surgeons are identified, satisfying the objective requirements of the follower. The algorithm performance is further improved via addition of different speed-up mechanisms such as the strengthened formulation of the lazy constraints, calculation of a dedicated lower bound, and a lazy constraint remembering mechanism. A comparison with a compact MIP formulation shows that the branch-and-price approach can find an optimal solution in an order of magnitude better time. Third, we obtain relevant managerial insights into the interaction between the surgeon head and individual surgeons based on the price of decentralisation and price of stability, analysing different appropriate scenarios for the surgeon scheduling problem under study.

\section{\label{section_problem}Problem description and formulation}
The problem under study encompasses planning and scheduling at the operational level, involving multiple agents who have to agree on a proper surgeon schedule and advance patient planning. In this decision-making process, we consider the surgeon head and the individual surgeons, whose roles conform to observation of real-life practice \citep{Johnston2019}. The problem under study focuses on the interplay between both parties, who are organised in a hierarchical manner. The surgeon head has the authority to determine both the allocation of OR capacity and advance patient planning, as long as the patient plan optimises the patient priority from the perspective of individual surgeons. In Section~\mbox{\ref{notation}}, we discuss the problem characteristics and context. Section~\mbox{\ref{interaction}} describes the interaction between the involved agents and the decisions taken by the different actors. Section~\mbox{\ref{subsection_formulation}} provides a mathematical formulation. 

\subsection{\label{notation}Characteristics and assumptions}

The problem regards a short-term horizon that comprises a set of days $D$ (index $d$) and a set of rooms $R$ (index $r$). We consider a particular surgeon group that involves a set of individual surgeons $S$ (index $s$) and a surgeon head, at the top of the group hierarchy. Each surgeon $s$ has a waiting list of elective patients $P_s$ (index $p$), whom the surgeon wants to include in the patient planning for the upcoming short-term horizon. The set of patients $P$ involves patients from all involved surgeons ($P = \bigcup_{s\in S} P_s$). The surgery of patients is characterised by a particular duration $\delta_p$. Discrete priority weights $\pi_{p}^{LP}$ and $\pi_p^{FP}$ are stipulated, indicating the preference to include patient $p$ in the patient planning as defined by the surgeon head and involved surgeon, respectively. A higher priority weight implies a higher preference for including the patient.

The total OR time or capacity allotted to the considered surgeon group is represented by $C$. Surgeons and patients are assigned to a set of so-called operational blocks $B$ (index $b$), which are uniquely defined by a particular start time $t$ and duration $\Gamma_b$, aligned to the tactical blocks specified in the MSS. The set $B_{d}$ represents the operational blocks on day $d$ ($B = \bigcup_{d\in D} B_{d}$). A block is composed of a set of consecutive time slots, each having a particular duration, which represents the granularity for scheduling patients. The possible block start times are designated by the set $T$ (index $t$), counted in time slots. As lengths of blocks can vary and need to be determined, blocks can overlap. The set $O_{dt} \subseteq B$ represents the set of blocks overlapping at time $t$ on day $d$. The number of blocks assigned to an individual surgeon is restricted by a maximum allowed number of blocks over the planning horizon ($v^h$) and per day ($v^d$). The former is imposed to limit the dispersion of blocks assigned to a particular surgeon over the planning horizon. The latter reflects the scheduling approach surgeons adopt for their planned patients, either performing all planned procedures consecutively in a single block ($v^d = 1$) or not. The set $N$ consists of tuples ($s$,$b$), representing blocks $b$ during which surgeon $s$ is unavailable to perform surgery. These unavailable blocks may either be allocated to a different surgeon group in the MSS or not aligned with the surgeons' preferences, which are passed to the surgeon head before the surgeon schedule is constructed. 

The structure of the leader and follower problems and their objective components are discussed in detail in Section \ref{interaction}. Below, the relevant assumptions and their implications that apply to the problem under study are stated:

\begin{itemize}[noitemsep]
\item[1.]	The problem under study is deterministic as all relevant parameters (e.g., surgery duration, priority weights) are assumed to be known \citep[e.g.][]{Roshanaei2017, Akbarzadeh2025}. 
\item[2.]	OR time and surgeons are explicitly considered for the planning of patients, as these are the most critical resources. Other resources (e.g., nursing staff) accommodate the schedule as required \citep[e.g.,][]{Roshanaei2020, Park2021}. 
\item[3.]	The activities in the OR department are organised following the block scheduling framework. Accordingly, we assume the existence of a long-term, cyclic MSS established through higher-level strategic and tactical decision-making \citep{Agnetis2014}. The MSS allocates so-called tactical blocks to specific surgeon groups, granting priority access to these surgeons during the time frames designated by the assigned blocks.
\item[4.]	The problem under study allocates operational blocks to individual surgeons, in accordance to empirical observations and previous studies \citep[e.g.,][]{Akbarzadeh2025}. Compared to tactical blocks specified in the MSS, these operational blocks may consist of a smaller number of time slots. The defined blocks in set $B$ ensure their start times and durations are consistent, preventing them from finishing later than the end times of the associated tactical blocks. The length of operational blocks can vary in correspondence to the volume and surgery duration of planned patients, possibly accounting for a minimum block length and fixed step size. For example, blocks may have a length ranging between two hours and an entire OR day of eight hours incrementing in steps of two hours. Assuming time slots have a duration of 15 minutes, this results in block sizes of 8, 16, 24, or 32 time slots, which is illustrated in Figure \ref{fig:configurations}. The possible block start times are designated by $t \in T = \{0, 8, 16, 24\}$.

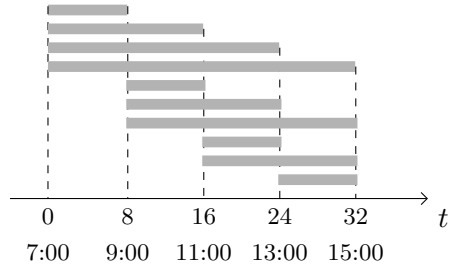
\begin{figure}
    \centering
    \begin{tikzpicture}[scale=1.0]
        \draw[-{Straight Barb}] (0.0,-2.50)--(5.5,-2.50) node[anchor=north west]{$t$};

        \node[rectangle, draw=gray!60, fill=gray!60, very thick, minimum height=1mm, minimum width=10mm, inner sep=0mm, anchor=west] at (0.5, 0) () {};
        \draw[dashed] (0.5,0.05)--(0.5,-2.50) node[anchor=north, text width=1cm, align=center]{\footnotesize 0 \\ 7:00};
        \draw[dashed] (1.55,0.05)--(1.55,-2.50) node[anchor=north, text width=1cm, align=center]{\footnotesize 8 \\ 9:00};

        \node[rectangle, draw=gray!60, fill=gray!60, very thick, minimum height=1mm, minimum width=20mm, inner sep=0mm, anchor=west] at (0.5, -0.25) () {};
        \draw[dashed] (2.55,-0.25)--(2.55,-2.50) node[anchor=north, text width=1cm, align=center]{\footnotesize 16 \\ 11:00};

        \node[rectangle, draw=gray!60, fill=gray!60, very thick, minimum height=1mm, minimum width=30mm, inner sep=0mm, anchor=west] at (0.5, -0.5) () {};
        \draw[dashed] (3.55,-0.5)--(3.55,-2.50) node[anchor=north, text width=1cm, align=center]{\footnotesize 24 \\ 13:00};

        \node[rectangle, draw=gray!60, fill=gray!60, very thick, minimum height=1mm, minimum width=40mm, inner sep=0mm, anchor=west] at (0.5, -0.75) () {};
        \draw[dashed] (4.55,-0.75)--(4.55,-2.50) node[anchor=north, text width=1cm, align=center]{\footnotesize 32 \\ 15:00};

        \node[rectangle, draw=gray!60, fill=gray!60, very thick, minimum height=1mm, minimum width=10mm, inner sep=0mm, anchor=west] at (1.53, -1.0) () {};        

        \node[rectangle, draw=gray!60, fill=gray!60, very thick, minimum height=1mm, minimum width=20mm, inner sep=0mm, anchor=west] at (1.53, -1.25) () {};   

        \node[rectangle, draw=gray!60, fill=gray!60, very thick, minimum height=1mm, minimum width=30mm, inner sep=0mm, anchor=west] at (1.53, -1.5) () {};   

        \node[rectangle, draw=gray!60, fill=gray!60, very thick, minimum height=1mm, minimum width=10mm, inner sep=0mm, anchor=west] at (2.53, -1.75) () {}; 

        \node[rectangle, draw=gray!60, fill=gray!60, very thick, minimum height=1mm, minimum width=20mm, inner sep=0mm, anchor=west] at (2.53, -2.0) () {}; 

        \node[rectangle, draw=gray!60, fill=gray!60, very thick, minimum height=1mm, minimum width=10mm, inner sep=0mm, anchor=west] at (3.53, -2.25) () {}; 
        
    \end{tikzpicture}
    \caption{
        \textcolor{black}{Illustration of the set of possible blocks $B_d$ in a single room on day $d$, with each block having a different start time and/or duration}.
    }
    \label{fig:configurations}
\end{figure}
\item[5.]	\textcolor{black}{Surgeons can express time preferences via the set $N$ \citep[e.g.,][]{Penn2017}. Rooms that are available to the surgeon group, are assumed to be identical \citep[e.g.,][]{Guido2017}.} 
\item[6.]	The defined problem can be applied across various scheduling approaches adopted by surgeons to treat their planned patients.  When the maximum number of blocks ($v^d$) equals one, it is assumed that surgeons perform all surgeries planned on that day consecutively in the same operating room\textcolor{black}{, which implies that surgeons cannot be allocated to parallel, overlapping blocks in different rooms}. This practice is known to improve the OR utilisation as it streamlines the organisation of planned surgeries and avoids transition disruptions when surgeons switch rooms \citep[e.g.,][]{Cardoen2009,Wang2014}. When $v^d$ is larger than one, surgeons possibly may have breaks in between scheduled blocks or are alternating between rooms to manage surgeries in parallel. The latter is possible when the prevailing MSS assigns multiple blocks in parallel rooms to the group the surgeon \textcolor{black}{belongs} \citep[e.g.,][]{Agnetis2014, Zhang2015}. The choice between both approaches depends upon the surgeon preferences, the type and complexity of surgeries, and the institutional policy.
\item[7.]	We do not explicitly consider the setup times between a surgeon’s procedures. However, we assume these setup times are fixed and incorporated into the standard surgery duration $\delta_p$ used for advance patient planning \citep[see, e.g.,][]{Samudra2017,Kroer2018,Marques2019}.
\item[8.]	The advance planning of patients is guided by patient priority weights $\pi_{p}^{LP}$ and $\pi_p^{FP}$, which may be determined differently. Defined priorities are assumed to be calculated as a weighted sum of patient-specific considerations possibly related to the patient's disease type, health status, medical urgency, time spent on the waiting lists of the surgeons, marginal revenues, etc. \citep[see, e.g.,][]{Testi2008, Testi2009, Shah2009, Guda2016, Guido2017}, with the assumption that these can be aggregated and represented as a discrete weight. 
\end{itemize}

\subsection{\label{interaction}Bilevel multi-agent scheduling}
The \textit{leader problem} involves the decisions, objectives, and constraints of the surgeon head. The surgeon head assigns operational blocks $b \in B$ having specific lengths to individual surgeons and determines thus the allotted capacity and schedule for individual surgeons. The upper-level binary decision variables $y_{sb}$ equal 1 if block $b$ is assigned to surgeon $s$, and are 0 otherwise. The surgeon schedule is composed while accounting for different block scheduling constraints. The objectives of the surgeon head are to maximise utilisation of the allotted OR time to his/her group and weighted case volume treated, accounting for the priority of individual cases from his/her perspective. Maximising the OR utilisation corresponds to minimising the idle time of OR capacity allocated to the surgeon group. The idle time is the positive difference between the OR capacity $C$ and the duration of patients scheduled by individual surgeons. The case volume treated is optimised by minimising the penalty $\pi_{p}^{LP}$ associated with patients that have not been included in the advance patient planning. The surgeon head accounts for these two measures of performance with $\alpha$~and $\beta$ as respective weights. Note that both these objectives are dependent on the decisions taken by individual surgeons related to the planning of patients, such that it is essential for the leader that the derived solution is an equilibrium solution that cannot be deteriorated due to the behaviour of individual surgeons. 

In the \textit{follower problem}, each individual surgeon constructs independently an advance patient planning. Patients are planned via patient-to-block assignments relying on the lower-level decision variables $x_{pb}$, which are equal to 1 if patient $p$ is assigned to block $b$, stipulating the day and time frame patients are operated, and 0 otherwise. Patient planning decisions should align with the capacity allocated by the leader. The objective of individual surgeons is to maximise the reward associated with the patients scheduled, expressed via the patient priorities $\pi_p^{FP}$. As these priorities are non-negative, individual surgeons will maximise the number of surgeries scheduled taking these patient weights into account.

Both hierarchical decision problems optimise the priority of treated patients. The priorities set by the two parties can be different from each other. The patient priority is the only interest of individual surgeons and an agreement with the surgeon head can only be made when there is no possibility of composing a surgical case planning with a higher reward. The bilevel optimisation problem under study ensures that, given the hierarchical decisions taken by the leader, the advance patient planning is optimal from the perspective of individual surgeons such that they will not change the outcome for the leader. Accordingly, the problem can be seen as a non-cooperative game assuming a leader and several followers. Though, the problem implicitly relies on the optimistic assumption \citep{Kalashnikov2016}, i.e., when the follower problem possesses alternative patient plans with an optimal priority, then the plan will be selected that is the most favourable for the leader, optimising the objective of the leader. We assume that the leader has perfect knowledge about these priorities set by the followers. When one studies the game from the leader's point of view, it is a Stackelberg game where the leader offers OR time slots, and followers assign their patients to the slots~\citep{Roughgarden2004}. Then, the problem is to determine the way the leader should allocate the OR time slots such that the leader optimises his objective and the entire solution is a Nash equilibrium, i.e., no follower can improve its objective function without taking OR time from other surgeons.

\begin{figure}[!ht]
\begin{center}
\includegraphics[width=6.5in]{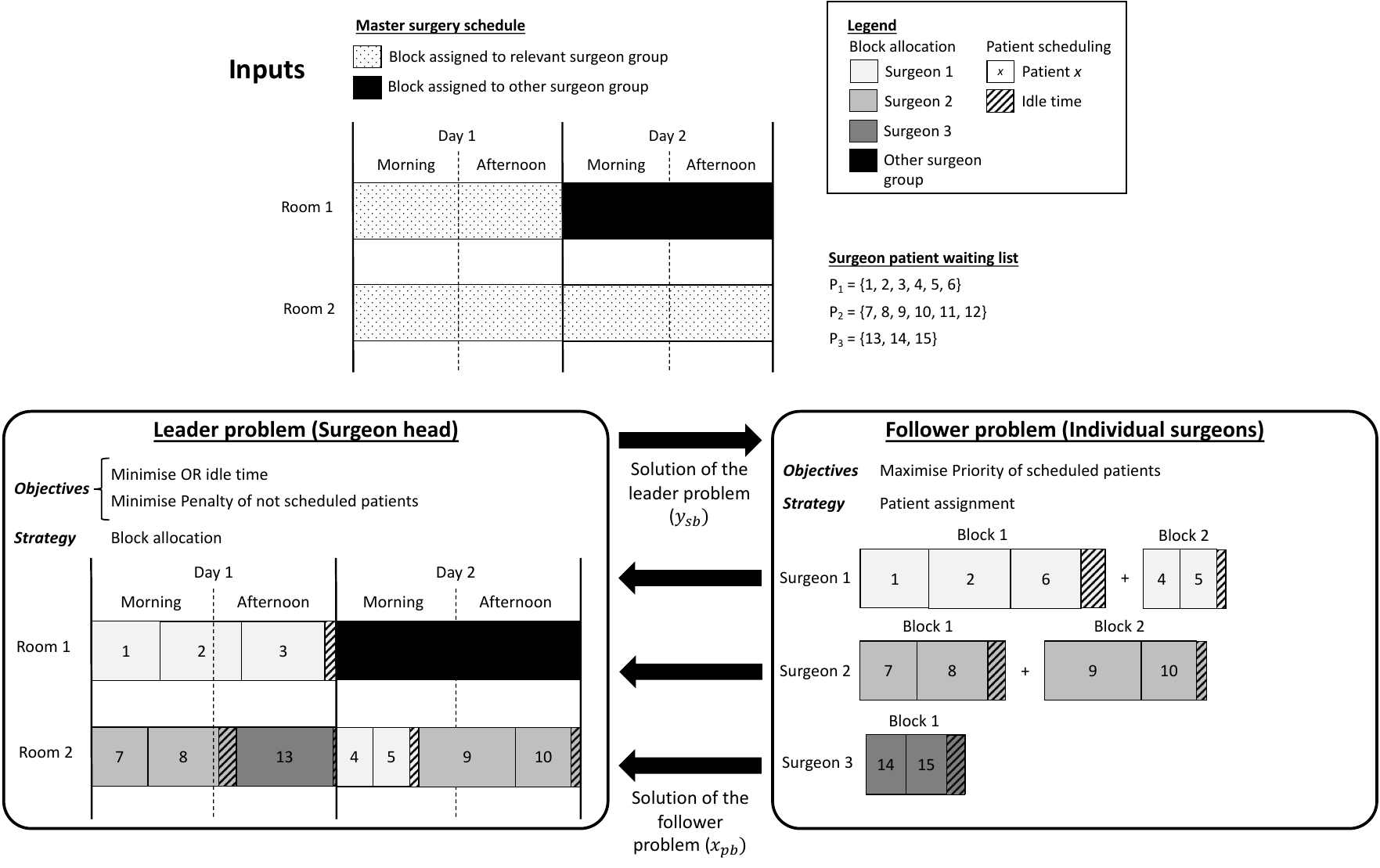} 
\caption{Problem context of the bilevel surgeon scheduling problem}
\label{Figure_Problem_Context}
\end{center}
\end{figure}

Figure \ref{Figure_Problem_Context} illustrates the problem under study. The example assumes a planning horizon of two days and two ORs. The figure displays the input MSS and patient waiting lists of the individual surgeons. The capacity allocated to the surgeon group embodies day 1 in room 1 and days 1 and 2 in room 2. The surgeon group involves three surgeons 1, 2, and 3, who have patients 1 to 6, patients 7 to 12, and patients 13 to 15 on their respective patient waiting lists. The solution yielded via solving the leader problem (see the schedule on the left in Figure \ref{Figure_Problem_Context}) allocates two blocks to surgeon 1 (on day 1 in room 1 and day 2 in room 2) and surgeon 2 (on day 1 in room 2 and day 2 in room 2), and one block to surgeon 3 (on day 1 in room 2). The patient schedule derived by the individual surgeons (follower problem) does not comply with the objective of the leader. More specifically, the patient schedule composed by the individual surgeons (see the schedule on the right in Figure \ref{Figure_Problem_Context}) shows a larger amount of idle OR time (hatched areas) and potentially a different penalty for omitting specific patients in the schedule. The latter stems from surgeons 1 and 3 who allocated different patients in their schedule according to postulated patient priorities, i.e., patient 6 instead of patient 3 (surgeon 1) and patients 14 and 15 instead of patient 13 (surgeon 3). As a result, the presented leader solution is not bilevel feasible. On the other hand, the solution constructed by individual surgeons is bilevel feasible, but it may not be optimal from the leader's point of view. Therefore, the schedule should be constructed by the leader, but needs to respect the objectives of surgeons.


\subsection{Mathematical problem formulation} 
\label{subsection_formulation}
The bilevel problem formulation for the proposed multi-agent surgeon scheduling problem is presented below.

\textbf{Notation}
\vspace{-2mm}
\footnotesize
\begin{longtable}{lp{0.9 \textwidth}}
	\multicolumn{2}{l}{{\textit{Sets}}}\\	
	\hspace{0.2cm}$S$ &  The set of surgeons ($S = \{0,1,\ldots, |S|-1\}$, with index $s$)\\
    \hspace{0.2cm}$P$ &  The set of patients ($P = \{0,1,\ldots, |P|-1\}$, with index $p$)\\
    \hspace{0.2cm}$P_{s}$ & The set of patients handled by surgeon $s$\\
	\hspace{0.2cm}$R$ &  The set of operating rooms ($R = \{0,1,\ldots, |R|-1\}$, with index $r$)\\
    \hspace{0.2cm}$D$ &  The set of days ($D = \{0,1,\ldots, |D|-1\}$, with index $d$)\\
    \hspace{0.2cm}$B$ &  The set of operational blocks ($B = \{0,1,\ldots, |B|-1\}$, with index $b$)\\
    \hspace{0.2cm}$B_d$ &  The set of operational blocks on day $d$ ($B_d = \{0,1,\ldots, |B_d|-1\} \subseteq B$, with index $b$)\\
    \hspace{0.2cm}$T$ & The set of possible start times of operational blocks ($T = \{0,1,\ldots, |T|-1\}$, with index $t$)\\
    \hspace{0.2cm}$O_{dt}$ & The set of blocks overlapping at time $t$ on day $d$\\
    \hspace{0.2cm}$N$ & The set of tuples $(s,b)$, representing the blocks $b$ during which surgeon $s$ is unavailable to perform surgery\\
	\multicolumn{2}{l}{\textit{Parameters}}\\
    \hspace{0.2cm}$\delta_{p}$ & The surgery duration of patient \textit{p}\\	
    \hspace{0.2cm}$C$ & The capacity allotted to the involved surgeon group\\
    \hspace{0.2cm}$\Gamma_b$ & The duration of block $b$\\
    \hspace{0.2cm}$v^h$ & The maximum allowed number of blocks allocated to a surgeon over the planning horizon \\
    \hspace{0.2cm}$v^d$ & The maximum allowed number of blocks allocated to a surgeon per day \\
	\hspace{0.2cm}\textcolor{black}{$\pi_{p}^{LP}$} & The priority weight associated with the surgery of patient $p$, stipulated by the surgeon head \textcolor{black}{(LP: Leader Problem)}\\
    \hspace{0.2cm}\textcolor{black}{$\pi_p^{FP}$} & The priority weight associated with the surgery of patient $p$, stipulated by the operating surgeon \textcolor{black}{(FP: Follower Problem)}\\
    \hspace{0.2cm}$\alpha$ & Objective weight associated with minimisation of the idle time, stipulated by the surgeon head\\
    \hspace{0.2cm}$\beta$ & Objective weight associated with minimisation of the priority weights of patients, not included in the advance planning, as defined by the surgeon head\\
	\multicolumn{2}{c}{}\\
	\multicolumn{2}{l}{{\textit{Decision variables}}}\\
	\hspace{0.2cm}$y_{sb}^{}$ & 1, if block $b$ is allocated to surgeon \textit{s}; 0, otherwise\\
    \hspace{0.2cm}$x_{pb}^{}$ & 1, if patient \textit{p} is allocated to block \textit{b}; 0, otherwise\\
	\hspace{0.2cm}$F$ & Objective value for the surgeon head or leader\\
	 \hspace{0.2cm}$f_s$ & Objective value for surgeon $s$ or follower
	\label{tab:CompactNotation}
  \label{tab:Compactlabel}%
\end{longtable}%
\addtocounter{table}{-1}
\normalsize

\vspace{-2mm}
\textbf{Mathematical formulation}\\
\vspace{-2mm}
\footnotesize
\begin{flalign}
\textrm{Min } F = \alpha(C - \sum_{p \in P}\sum_{b \in B} \delta_p x_{pb}) + \beta\sum_{p\in P} \pi_{p}^{LP} (1-\sum_{b \in B} x_{pb}) && \label{eq:obj_leader}
\end{flalign}
\begin{flalign}
 \textrm{s.t.} \;     & \sum_{b\in O_{dt}} \sum_{s\in S} y_{sb} \leq |R| & \forall d\in D,\ \forall t \in T \label{eq:2} \\
   & \sum_{b\in B_{d}} y_{sb} \leq v^d & \forall s\in S,\ \forall d\in D \label{eq:3} \\
   & \sum_{b\in B} y_{sb} \leq v^h & \forall s\in S \label{eq:4}
   \end{flalign}
   \vspace{-4mm}
\begin{flalign}
  \quad       & \textcolor{black}{y_{sb} \in \begin{cases}
           \{0, 1\}  \hspace{10cm}&  \hspace{9cm}\hfill \forall (s,b) \in (S\times B)\setminus N\\
           \{0\}  & \hfill \makebox{$\forall (s,b) \in N$} 
       \end{cases}} \label{eq:dom_leader}   
\end{flalign}
   \vspace{-4mm}
\begin{flalign}
 \quad \; \;  & x_{pb} \in argmax\{f_s = \sum_{p\in P_s} \pi_p^{FP}  \sum_{b\in B} x_{pb}| (\ref{eq:7})-(\ref{eq:dom_follower})\} & \forall s\in S \label{eq:obj_follower} \\
    & \textrm{s.t.} \;  \; \; \; \, \sum_{b\in B} x_{pb} \leq 1 & \forall p\in P \label{eq:7}\\
    & \qquad \; \, \sum_{p \in P_s} \delta_p x_{pb} \leq \Gamma_b y_{sb} & \forall s \in S, b\in B \label{eq:8}\\
    & \qquad \; \, x_{pb} \in \{0, 1\} &  \forall p \in P, \forall b \in B &  \label{eq:dom_follower} 
\end{flalign}
\normalsize

Objective (\ref{eq:obj_leader}) represents the objective function of the surgeon head, which contains lower-level decision variables. This entity aims to minimise the idle time in the blocks allotted to the surgeon group, which corresponds to maximising the OR utilisation, and to minimise the priority penalty associated with patients that are not included in the patient planning. Both these objectives are dependent on the patient planning of individual surgeons, which is determined by the follower problem. \textcolor{black}{Note that objective (\ref{eq:obj_leader}) can be rewritten to an equivalent form, i.e., $\textrm{Max} \; \alpha \sum_p \sum_b \delta_p \; x_{pb} + \beta \sum_p \sum_b \pi^{LP}_p \; x_{pb} - cte$ (with $cte$ indicating a constant value). This formulation shows that the objective maximises a weighted combination of the total planned surgery duration and the priority weights of the scheduled patients, minus a constant term.} \textcolor{black}{Constraint (\ref{eq:2}) limits the number of operational blocks in progress at any time $t$ on day $d$ to the number of available rooms. As rooms available to the considered surgeon group are assumed to be identical, blocks are not linked to specific rooms. This allows to formulate the constraint relative to concurrent block assignments in an aggregated manner. Without this constraint, more blocks than available rooms may be selected that are overlapping, which would be infeasible. Note that this constraint also implicitly imposes that the allotted blocks align with the available capacity per room owing to the definition of the set of blocks $B$.} Constraints (\ref{eq:3}) and (\ref{eq:4}) restrict the number of blocks that can be assigned to an individual surgeon per day and over the entire planning horizon, respectively. \textcolor{black}{Parameter $v^d$ in constraint (\ref{eq:3}) controls the scheduling approach adopted by surgeons to treat planned patients (cf. Assumption 6).} Constraint (\ref{eq:dom_leader}) defines the binary domain of the upper-level variables. \textcolor{black}{Note that this domain constraint also stipulates that a surgeon $s$ cannot be assigned to a block $b$, as the block may be assigned to a different surgeon group according to the MSS or the individual surgeon is not available.} \\
The bilevel problem further includes the lower-level follower problem in the constraints of the leader problem, which basically embodies multiple knapsack problems and is strongly NP-hard. Equation (\ref{eq:obj_follower}) indicates that a feasible solution for the leader problem should be an optimal solution for the follower problem given their objective $f_s$ ($\forall s \in S$), which maximises the patient priority from the perspective of the individual surgeons.
The feasible region of the follower problem regards constraints (\ref{eq:7})-(\ref{eq:dom_follower}). Constraints (\ref{eq:7}) imposes that each patient can be scheduled at most once. Constraint (\ref{eq:8}) ensures that the duration of scheduled surgical cases does not exceed the capacity of block $b$ assigned to surgeon $s$. In addition, the constraint models the relationship between leader and followers decisions, i.e., if a patient is scheduled in a block, the involved surgeon needs to be assigned to the block. Constraint (\ref{eq:8}) connects upper and lower-level decision variables. The domain of the variables related to the follower problem is stated in constraints (\ref{eq:dom_follower}). \\
A solution is called bilevel feasible if it satisfies all restrictions (\ref{eq:2})-(\ref{eq:dom_follower}). The feasible region that contains all bilevel feasible solutions is called the inducible region. These solution points embody an optimal follower decision, i.e., a maximised patient priority in response to the leader decision. A bilevel optimal solution is bilevel feasible and optimises the leader's objective $F$ (eq. (\ref{eq:obj_leader})).


\section{\label{section_methodology} Lazy constraint generation approach}

The bilevel surgeon scheduling and surgery planning problem formulated in Section~\ref{subsection_formulation} cannot be stated as a single MIP based on duality theory as the decision variables of the follower problem are binary and not continuous. \textcolor{black}{Relevant techniques for solving bilevel optimization problems involving follower problems with continuous and binary variables can be found in the review papers of \cite{Colson2007} and \cite{Kleinert2021}, respectively.} The proposed solution technique considers separate formulations of the leader and follower problems and searches through the bilevel search space by continuously adding \textit{lazy constraints} restricting the region such that the leader has to respect the objective requirements of the follower. This means that whenever the MIP solver, solving the leader problem, finds an integer feasible solution, our approach tests whether this solution is bilevel feasible in a \textit{callback} function that solves the follower problem for every surgeon. If the callback function proves optimality of patient assignments for every surgeon, the solution is bilevel feasible. In that case, the solver has found a new candidate solution. If the yielded integer solution is not bilevel feasible, lazy constraints are formulated based upon the follower outcomes of the callback function, connecting the allocation of OR blocks (leader decision) with the patient planning accounting for the priorities of the individual surgeons (follower responses) to incorporate information related to the bilevel feasibility of integer solutions. These lazy constraints are added to the leader problem formulation, which cuts away the bilevel infeasible solution. In both cases, the MIP solver continues to search for a different solution for the leader problem. The MIP formulations for the leader problem and the follower problem are discussed in detail in Section~\ref{subsection_leader_follower}. The modelling of lazy constraints is explained in Section~\ref{subsec:LC}.

\subsection{\label{subsection_leader_follower}Formulation of leader and follower problem}

The leader problem is formed by equations (\ref{eq:2})-(\ref{eq:dom_leader}) and (\ref{eq:7})-(\ref{eq:dom_follower}), omitting the follower objective constraint (\ref{eq:obj_follower}). The callback formulates a follower problem for every surgeon $s \in S$, which corresponds to equations (\ref{eq:obj_follower})-(\ref{eq:dom_follower}). The follower problem for surgeon $s$ is defined as

\vspace{-3mm}
\footnotesize
\begin{flalign}
\textrm{Max } &{f}_s' = \sum_{p\in P_s} \pi_p^{FP}  \sum_{b\in B} x_{pb}' & \label{eq:obj_follower_callback}\\
 \textrm{s.t.}   & \sum_{b\in B} x_{pb}' \leq 1 & \forall p\in P_s \label{eq:7b}\\
    & \sum_{p \in P_s} \delta_p x_{pb}' \leq \Gamma_b y_{sb} & \forall b\in B \label{eq:8b}\\
    & x_{pb}' \in \{0, 1\}. &  \forall p \in P_s, \forall b \in B &  \label{eq:dom_follower_b}   
\end{flalign}
\normalsize

Variable $x_{pb}'$ has the same meaning as $x_{pb}$ as in the leader problem, but in the follower problem, these variables represent the decisions of surgeon $s$. On the other hand, the solution of the leader problem \textcolor{black}{determines values for the variables $y_{sb}$ and $x_{pb}$. The latter are used to calculate $f_s$. Both} $y_{sb}$ and $f_s$ are inputs to the callback, and thus are constants. While $y_{sb}$ is directly used in the follower problem in constraint (\ref{eq:8b}) and represents the leader's decision, $f_s$ is not used in the above-stated follower problem, but it is used later (cf. Section~\ref{subsec:LC}). The objective function of the follower problem is ${f}_s' = \sum_{p\in P_s} \pi_p^{FP}  \sum_{b\in B} x_{pb}'$, and follows from equation (\ref{eq:obj_follower}) defining the follower's behaviour.

When the leader problem finds a feasible integer solution, then the callback takes its solution, i.e., $y_{sb}$ and $f_s$, as input and, for every surgeon $s$, the callback formulates the follower problem. If for every surgeon $s \in S$ holds that ${f}_s'$ (computed from $x_{pb}'$) equals $f_s$ (computed from the leader's assignment $x_{pb}$), then the solution is bilevel feasible. Otherwise, the solution of the leader problem needs to be cut off by a lazy constraint. In this case, we define the parameters of the lazy constraint as a tuple $(s^c, y_{b}^c, x_{pb}^c, f^c)$, where $s^c$ is the follower, $y_{b}^c$ defines blocks assigned by the leader to surgeon $s^c$, i.e., $y_{b}^c = y_{{s^c}b}$, $x_{pb}^c$ are patients assigned in the follower problem, i.e., $x_{pb}^c = x_{pb}'$ ($\forall p \in P \setminus P_{s^c}:~x_{pb}^c = 0$), and $f^c$ is the optimal follower objective, i.e., $f^c = f_{s^c}'$. The upper index $c$ represents the counter of generated lazy constraints. Each time the follower problem is solved, this counter is increased by one. The reason to keep track of generated lazy constraints is that lazy constraints remain active inside the leader problem formulation and not only the last generated one.

\subsection{Lazy constraints and elimination of symmetries}
\label{subsec:LC}

Defining lazy constraints relying directly on $x_{pb}$ and $y_{sb}$ would be inefficient since \textcolor{black}{an equivalent block assignment - containing the same number of blocks of equal lengths but arranged with different timing - }can lead to the same solution from the leader's objective point of view, which is not cut off by a previously generated lazy constraint. \textcolor{black}{Because of the large degree of symmetry between possible block assignments, we define strengthened lazy constraints by introducing auxiliary binary variables $q_{slw}$ in the leader problem. These variables} equal 1 if $w$ blocks of length $l$ are assigned to surgeon $s$, and 0 otherwise. By defining $L$ as the set of all possible block lengths ($L = \{0,1,\ldots, |L|-1\}$, with index $l$) and $W_l$ as the set of all possible numbers of blocks of a given length $l$ that can be assigned to a single surgeon ($W_l = \{0,1,\ldots, |W_l|-1\}$, with index $w$), we can express the relationship between $y_{sb}$ and $q_{slw}$ in the leader problem via eqs. (\ref{eq:blocks1}) and (\ref{eq:blocks2}). 

\vspace{-4mm}
\footnotesize
\begin{flalign}
\qquad &\sum_{w \in W_l} w \; q_{slw} = \sum_{b \in B | \Gamma_b = l} y_{sb},  & \forall s \in S, \forall l \in L  \label{eq:blocks1} \\
&\sum_{w \in W_l} q_{slw} = 1. & \forall s \in S, \forall l \in L \label{eq:blocks2}
\end{flalign}
\normalsize

Based on these auxiliary variables, we define two types of lazy constraints, i.e.,

\textit{Objective-based Lazy Constraint} (\textit{O-LC})\\
The $O-LC$ connects an integer solution of the leader problem and a corresponding optimal solution of the follower problem by forcing the leader to assign patients such that $f_s$ is maximised if surgeon $s$ has a specific assignment of blocks. The $O-LC$ with counter index $c$ parametrised by tuple $(s^c, y_{b}^c, x_{pb}^c, f^c)$ is formulated as

\vspace{-4mm}
\footnotesize
\begin{flalign}
\qquad & \sum_{p \in P_{s^c}} \pi_p^{FP} \sum_{b \in B} x_{pb} + M \left( |L| - \sum_{l \in L} q_{{s^c}ln^c_{l}} \right) \geq f^c, & \label{ref:OBLC}
\end{flalign}
\normalsize

where $n^c_{l} = \sum_{b \in B | \Gamma_b = l} y_{b}^c$ is the number of blocks of duration $l$ assigned to surgeon $s^c$ specified by parameter $y_{b}^c$ in lazy constraint $c$. 
Parameter $M$ is a large positive value that enforces the logical condition determining whether the constraint is either relevant or redundant. A suitable value for $M$ is the sum of patient priorities for all patients relative to surgeon $s^c$, i.e., $\sum_{p\in P_{s^c}} \pi_p^{FP}$, as determined by the surgeon. The $O-LC$ imposes that whenever the leader assigns blocks defined by $n^c_{l}$, then $f_s \geq {f}^{c}$.

\textit{Assignment-based Lazy Constraint} (\textit{A-LC})\\
Unlike the previous type, the $A-LC$ does not directly target the follower objective but instead forces a specific assignment of patients. The $A-LC$ with counter index $c$ for surgeon $s^c$ is defined as

\vspace{-3mm}
\footnotesize
\begin{flalign}
\qquad & \sum_{p \in \mathit{PA^c}} \sum_{b \in B} x_{pb} - \sum_{p \in P_{s^c} \setminus \mathit{PA^c}} \sum_{b \in B} x_{pb} + M \left( |L| - \sum_{l \in L} q_{{s^c}ln^c_{l}} \right) \geq |\mathit{PA^c}|, & \label{ref:ABLC}
\end{flalign}
\normalsize

where $\mathit{PA^c} = \{ p \in P_{s^c} | x_{pb}^c = 1 \}$ is a set of patients assigned by the follower problem of surgeon $s^c$ for lazy constraint $c$. The first two terms on the left-hand side of the lazy constraint represent the sum of patient assignments with respect to $\mathit{PA^c}$, where assigned patients have positive signs and unassigned patients have negative signs. The third term is the same as in the case of $O-LC$ and turns the lazy constraint on and off according to variables $q_{{s^c}ln^c_{l}}$, referring to the number of blocks $n^c_{l}$ allocated with length $l$. The right-hand side of the lazy constraint is the number of assigned surgeries by the follower. Therefore, whenever the leader assigns blocks in accordance with $n^c_{l}$, the lazy constraint forces the leader problem to assign precisely the patients included in set $\mathit{PA^c}$. The motivation behind introducing $A-LC$ is that it should have a better linear relaxation compared to $O-LC$ and the ability to eliminate symmetric assignments of patients, which can be practical for problem instances assuming large sets of patients. For a computational comparison between both types, we refer to Section~\ref{section_results}.

Note that the use of $A-LC$ requires a modification of the follower problem. This aspect may not be obvious but it is essential to guarantee the optimality of the algorithm. This is stated in the following property:
\begin{property}
\label{prop:A-LC}
    The bilevel optimisation problem under study implicitly uses the optimistic assumption \citep{Kalashnikov2016}, i.e., when the follower problem possesses alternative optimal solutions, then one needs to select the surgery schedule that minimises the leader objective to ensure convergence. Correspondingly, the callback function for $A-LC$ must take into consideration the leader objective (\ref{eq:obj_leader}) and has to assume objective (\ref{eq:obj_follower_callback_LC-A}) instead of equation (\ref{eq:obj_follower_callback}) as the objective.
    
    \vspace{-6mm}
    \footnotesize
    \begin{flalign}    \label{eq:obj_follower_callback_LC-A}
    \textrm{Max } {f'}_s = \sum_{p\in P_s} \pi_p^{FP}  \sum_{b\in B} x_{pb}' - \frac{1}{M'} \left(
    \alpha \left( C - \sum_{p \in P_s}\sum_{b \in B} \delta_p x_{pb}' \right) + \beta\sum_{p\in P_s} \pi_{p}^{LP} \left(1-\sum_{b \in B} x_{pb}' \right) \right), && 
    \end{flalign}
    \normalsize
\end{property}
\begin{proof}
When the callback invokes the solving of a follower problem and selects the patients according to the follower objective, this assignment is fixed by lazy constraint (\ref{ref:ABLC}). However, for a given assignment of blocks to surgeon $s$, there may be two assignments of the patients, designated as $x_{pb}^{(1)}$ and $x_{pb}^{(2)}$ for $p \in P_s$, having the same follower objective $f_s$ but leading to a different contribution to the leader's objective, i.e., $F(x_{pb}^{(1)}) > F(x_{pb}^{(2)})$ assuming fixed decisions related to other surgeons. If the callback function returns $x_{pb}^{(1)}$, the $A-LC$ fixes this assignment and the leader problem formulation is not able to perform assignment $x_{pb}^{(2)}$. Consequently, the algorithm may miss the optimal solution.
\\
To fix this issue, the follower objective needs to be extended such that it can properly distinguish between assignments $x_{pb}^{(1)}$ and $x_{pb}^{(2)}$. Therefore, we incorporate the leader's objective as a secondary objective within the objective of the follower problem (\ref{eq:obj_follower_callback_LC-A}) solved in the callback procedure.  
Constant $M'$ in (\ref{eq:obj_follower_callback_LC-A}) is larger than the biggest value of the secondary objective. Assuming integer patient priorities $\pi_p^{FP}$, this will guarantee that the follower problem will generate a pattern for $x_{pb}^{(2)}$ and not for $x_{pb}^{(1)}$, which is more desired from the perspective of the leader.
\end{proof}

Accounting for this property, the validity of the defined lazy constraints can be stated, i.e.,

\begin{property}
Both $O-LC$ and $A-LC$ can substitute constraint (\ref{eq:obj_follower}) in the mathematical formulation of the multi-agent surgeon scheduling problem and represent valid inequalities.
\end{property}
\begin{proof}
Both lazy constraint $O-LC$ and $A-LC$ cut off only solutions that are not bilevel feasible, i.e., directly violate constraint (\ref{eq:obj_follower}). Lazy constraint $A-LC$ assigned patients such that $f_s$ is maximised, and thanks to the reformulation of the follower objective defined in Property~\ref{prop:A-LC} never cuts off an optimal solution. On the other hand, this reformulation of the follower objective is not required for $O-LC$ as this lazy constraint type does not define a specific assignment of patients but directly restricts $f_s$ such that it is maximal regarding blocks assigned to surgeon $s$. Therefore, both $O-LC$ and $A-LC$ contribute to the delineation of the inducible region of the problem.
\end{proof}

Disregarding which lazy constraint is used, there are two options to guarantee that the solution of the leader problem is bilevel feasible. Whenever the callback detects a bilevel infeasible solution, the first option is to add one lazy constraint for an arbitrary surgeon where it detects that ${f'}_s > f_s$. Another option is to add lazy constraints for all surgeons with these violations. The advantage of the first approach is that bilevel feasibility detection is faster as the callback usually does not need to solve all follower problems, while the advantage of the second one is that it prunes the solution space more.








\section{\label{section_BaP} Branch-and-price approach}

The above-introduced MIP formulation of the leader problem (cf. Section \ref{subsection_leader_follower}) extended by lazy constraints can find an optimal solution. Nevertheless, the computational complexity significantly grows with a rising number of patients and surgeons, as will be illustrated in Section~\ref{section_results}. To propose a solution methodology with better scalability, we suggest decomposing the problem using Dantzig Wolfe's decomposition and solving it using a branch-and-price approach dedicated to the problem under study. By this, we want to illustrate several important properties for decomposing a bilevel optimisation problem and show techniques to improve its performance.

\begin{figure}[htbp]
    \centering
    \includegraphics[width=1.0\linewidth]{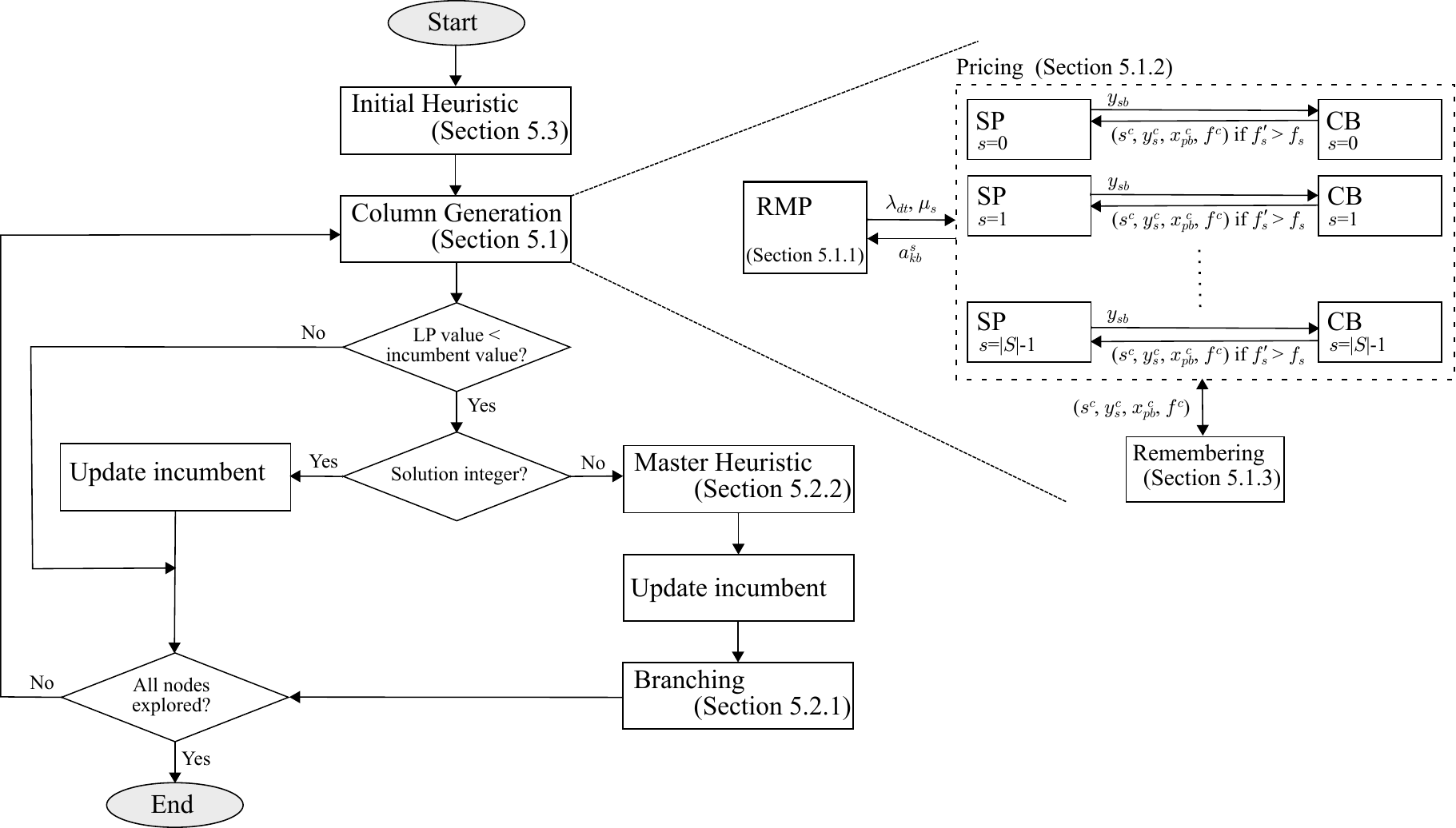}
    \caption{Flowchart of the proposed branch-and-price algorithm}
    \label{fig:column_generation}
\end{figure}

Branch-and-price is a hybrid optimisation approach for solving large-scale MIPs that combines a branch-and-bound search with a column generation procedure in each node of the search tree. In column generation, discussed in Section \ref{section_BaP_CG}, the original problem is reformulated via the Dantzig-Wolfe decomposition into a so-called master problem that ties smaller subproblems together. The master problem variables, also called pattern variables, are defined via subproblems having original constraints. In this way, the master problem combines solutions provided by subproblem(s) exploiting the decomposable structure of the original problem. Since column generation solves a linear relaxation of the original problem, the branch and price uses branch and bound to achieve an optimal integer solution, which is explained in Section \ref{section_BaP_BB}. The procedure is initialised using a heuristic procedure, described in Section \ref{section_BaP_IH}, to improve the early bounds employed for pruning the search space and speeding up convergence of column generation. The proposed branch-and-price algorithm for the studied bilevel optimization problem is illustrated in Figure~\ref{fig:column_generation} in the form of a flowchart and it is further described below. The figure highlights especially the column generation algorithm (see the right part of the figure), which tackles the bilevel aspect of the optimisation.

\subsection{\label{section_BaP_CG} Column generation}
Column generation iterates between a restricted master problem (RMP) and a subproblem (SP) that tries to identify columns having the potential to further improve the objective function value of the RMP. To guarantee that the optimal solution of the RMP corresponds to an optimal LPR solution of the problem under study, we try to identify suitable patterns by solving the subproblem for each surgeon. Suitable patterns encompass (bilevel) feasible surgeon schedules having a positive reduced cost. These columns are added to the column set that is considered in the RMP. If no suitable column can be identified, an optimal solution for the linear relaxation of the bilevel problem (\ref{eq:obj_leader})-(\ref{eq:dom_follower}) has been found.

\subsubsection{\label{section_BaP_MP} Restricted master problem}

The decomposition applied to the bilevel surgeon scheduling and surgery planning problem divides the problem based on the surgeons and the construction of their schedules. This is supported by the fact that each surgeon maintains his/her own waiting list of patients. The decomposition defines a \textit{schedule} of a surgeon, which is an assignment of blocks to the surgeon. The schedule restricts the number of blocks that can be assigned to a surgeon per day and over the planning horizon (see eqs.~(\ref{eq:3}) and (\ref{eq:4})). Moreover, the schedule is associated with a bilevel feasible assignment of patients, which is given by the surgeon's (follower) objective (see eq.~(\ref{eq:obj_follower})). The defined master problem does not use original variables $y_{sb}$ and $x_{pb}$ but relies on pattern variables $\theta_k^s$, which equal one if schedule $k$ is assigned to surgeon $s$ and zero otherwise. Schedule $k$ of surgeon $s$ is defined via vector $a_{kb}^s$, called a pattern. An element of the vector $a_{kb}^s$ equals one if block $b$ belongs to schedule $k$ and zero otherwise. Further, two scalars are associated with each pattern, i.e., $\Delta_k^s$ and $\rho_k^s$. The first, $\Delta_k^s$, represents the total duration of patient surgeries assigned by surgeon $s$, reflecting the contribution of schedule $k$ to the OR utilisation. The second, $\rho_k^s$, indicates the penalty associated with the patients whose surgery is not performed if schedule $k$ is selected. 

The master problem in this decomposition includes all feasible surgeon schedules $k \in K_s$ for every surgeon $s \in S$. In column generation, we typically do not assume all possible patterns due to their exponential number w.r.t. the instance size. Consequently, the procedure considers a RMP that contains only a subset of these pattern variables $\bar K_s \subset K_s$ and relaxes the binary domain conditions, i.e., $\theta_k^s \in [0,1]$. This relaxed form of the RMP is formulated by Equations (\ref{eq:obj_master})-(\ref{eq:master_var}). The objective function (\ref{eq:obj_master}) is a reformulation of the leader objective function penalising the underutilisation of ORs and patients waiting for surgery.  The RMP has two types of constraints. Constraints (\ref{eq:master_capacity}) restrict the number of overlapping blocks based on the number of ORs, and constraints (\ref{eq:master_pattern}) guarantee that exactly one pattern is selected for every surgeon. The included patterns are generated at run time by solving subproblems. Note that the initial patterns that are needed for the first run of column generation, i.e., before any subproblem has been solved, are generated in a trivial manner stipulating that every surgeon has no block assignment, which leads to a feasible solution to the RMP. 

\vspace{-3mm}
\footnotesize
\begin{flalign}
\textrm{Min } F = \alpha(C - \sum_{s \in S}\sum_{k \in \bar K_s} \Delta_k^s \theta_k^s) + \beta\sum_{s\in S}\sum_{k\in \bar K_s} \rho_k^s \theta_k^s && \label{eq:obj_master}
\end{flalign}
\begin{flalign}
  \textrm{s.t.} \; & \sum_{b\in O_{dt}} \sum_{s\in S} \sum_{k\in \bar K_s} a_{kb}^s \theta_k^s \leq |R| & \forall d\in D,\ \forall t \in T \label{eq:master_capacity} \\
   & \sum_{k \in \bar K_s} \theta_k^s = 1 & \forall s \in S \label{eq:master_pattern} \\
   & \theta_k^s \geq 0 & \forall s \in S,\ \forall k \in \bar K_s \label{eq:master_var}
\end{flalign}
\normalsize

\subsubsection{\label{section_BaP_SP} Subproblem}
The surgeon schedules represented by patterns $a_{kb}^s$, which are relevant for the RMP, are generated via solving the subproblem. Since it is hard to search for suitable patterns in the primal RMP formulation, the subproblem uses the dual formulation of the RMP, where variables are transformed into corresponding dual constraints. This means that the subproblem does not search for a missing primal variable $\theta_k^s$ but a pattern for which the corresponding dual constraint is violated. Since the RMP relies only on variables $\theta_k^s$, the dual RMP formulation is formed by a single set of constraints, i.e., 

\vspace{-3mm}
\footnotesize
\begin{flalign}
    \sum_{d \in D} \sum_{t \in T} \sum_{b \in O_{dt}} a_{kb}^s \lambda_{dt} + \mu_s \leq \beta \rho_k^s - \alpha \Delta_k^s, && \forall s \in S, \forall k \in \bar K_s
    \label{eq:dual_constraint}
\end{flalign}
\normalsize
where $\lambda_{dt} \leq 0$ and \textcolor{black}{$\mu_s \in \mathbb{R}$} are dual variables related to constraints (\ref{eq:master_capacity}) and (\ref{eq:master_pattern}), respectively. The subproblem searches a pattern $a_{kb}^s$ for which the corresponding dual constraint is infeasible, i.e., $\sum_{b \in O_{dt}} a_{kb}^s \lambda_{dt} + \mu_s > \beta \rho_k^s - \alpha \Delta_k^s$ and it is formulated by equations (\ref{eq:obj_subproblem})-(\ref{eq:sub_y_follower}) considering surgeon $s \in S$. Accordingly, the subproblem relies on variables $x_{pb}$ ($p \in P_s$) and $y_{sb}$, considering only those assignments involving surgeon $s$. Objective (\ref{eq:obj_subproblem}) maximises the reduced cost for given dual prices $\lambda_{dt}$ and $\mu_s$ stemming from solving the RMP. The constraints (\ref{eq:sub_3})-(\ref{eq:sub_y_follower}) of the subproblem define a feasible schedule of surgeon $s$ in the same way as it is described in Section~\ref{subsection_formulation} with addition of constraints (\ref{eq:blocks1}) and (\ref{eq:blocks2}) defining the auxiliary variables $q_{slw}$ needed for the lazy-constraint generation. 

\vspace{-3mm}
\footnotesize
\begin{flalign}
\textrm{Max } \sum_{d \in D} \sum_{t \in T} \sum_{b \in O_{dt}} \lambda_{dt} y_{sb} + \mu_s - \beta \rho_s + \alpha \Delta_s   &&
\label{eq:obj_subproblem}
\end{flalign}
\begin{flalign}
 \textrm{s.t.} \; \;  & \sum_{b\in B_{d}} y_{sb} \leq v^d & \forall d\in D \label{eq:sub_3} \\
   & \sum_{b\in B} y_{sb} \leq v^h &  \label{eq:sub_4} \\
   & \sum_{b\in B} x_{pb} \leq 1 & \forall p\in P_s \label{eq:sub_7}\\
   & \sum_{p \in P_s} \delta_p x_{pb} \leq \Gamma_b y_{sb} & b\in B \label{eq:sub_8}\\
   & \sum_{w \in W_l} w \; q_{slw} = \sum_{b \in B | \Gamma_b = l} y_{sb}  & \forall l \in L \label{eq:sub_blocks1} \\
   & \sum_{w \in W_l} q_{slw} = 1 & \forall l \in L \label{eq:sub_blocks2}\\
   & \rho_s = \sum_{p\in P_s} \pi_{p}^{LP} (1-\sum_{b \in B} x_{pb}) \\
   & \Delta_s = \sum_{p \in P_s}\sum_{b \in B} \delta_p x_{pb}\\
 & x_{pb} \in \{0, 1\} & \forall p \in P_s, \forall b \in B & \label{eq:sub_x_follower}
 \end{flalign}
   \vspace{-4mm}
\begin{flalign}
\quad    \;   & \textcolor{black}{y_{sb} \in \begin{cases}
           \{0, 1\}  \hspace{10cm}&  \hspace{8.8cm}\hfill \forall (s,b) \in (S\times B)\setminus N\\
           \{0\}  & \hfill \makebox{$\forall (s,b) \in N$} 
       \end{cases}}  \label{eq:sub_y_follower}   
\end{flalign}
\normalsize

As discussed in Section \ref{section_methodology}, yielded surgeon schedules should be bilevel feasible. Correspondingly, the subproblem applies a lazy-constraint generation approach to guarantee this property. Thanks to the applied decomposition, it is not necessary to check bilevel feasibility for all surgeons because the subproblem concerns a single surgeon only. Hence, whenever the subproblem finds a feasible (integer) solution, its bilevel feasibility is checked by the callback problem (\ref{eq:obj_follower_callback})-(\ref{eq:dom_follower_b}), and when it is not bilevel feasible, a lazy constraint, i.e., $O-LC$ or $A-LC$, is added to the subproblem and the search for the bilevel feasible surgeon schedule with the highest reduced cost is continued. The interaction between subproblems and callbacks is illustrated in Figure~\ref{fig:column_generation}. Similar to the note made in Section~\ref{subsec:LC}, the use of $A-LC$ in the subproblem requires a modification of the callback problem. To distinguish between two assignments of patients that are equally good for the follower but contribute differently to the leader's objective, we add a secondary objective reflecting the leader objective to the objective of the callback problem, which is then stated as

\vspace{-4mm}
\footnotesize
\begin{flalign}
\textrm{Max } {f'}_s = \sum_{p\in P_s} \pi_p^{FP}  \sum_{b\in B} x_{pb}' - \frac{1}{M'} \left(- \alpha \Delta_s + \beta \rho_s \right),&& 
\end{flalign}
\normalsize
where $M'$ is larger than the biggest value of $\left( -\alpha \Delta_s + \beta \rho_s \right)$.

The problem decomposition that we use to solve the studied problem provides a property that is crucial for efficient problem solving, as will be illustrated later in Section~\ref{subsection_benchmark}.

\begin{property}
    A decomposed problem formulation that integrates the required bilevel feasibility in the defined subproblem, ensures a proper linear relaxation of the inducible region of the bilevel optimisation problem and, hence, provides a much tighter bound in comparison to the linear relaxation of the monolithic formulation.
\end{property}
\begin{proof}
    Apart from the fact that Dantzig Wolfe's decomposition provides a tighter lower bound compared to the linear relaxation of the original MIP \citep{Vanderbeck2010}, the columns generated in the branch-and-price method represent bilevel feasible patterns. This cannot be assured by the linear relaxation of the MIP from Section~\ref{section_methodology} as the leader problem is actually a relaxation of the true bilevel optimisation problem. Since a monolithic formulation cannot identify bilevel feasible solutions in advance, which need to be identified via callbacks, a decomposed formulation, where the subproblem inherently ensures bilevel feasible solutions, provides a tighter bound.    
\end{proof}

\subsubsection{\label{section_BaP_SM} Speed-up mechanisms}
To speed up the solving of the subproblem, searching for a surgeon schedule that is bilevel feasible, we consider the technique of \textit{lazy-constraint remembering} (LCR). This procedure is based on the following property:
\begin{property}
A lazy constraint generated in any run of the subproblem for surgeon $s$ is valid at any other run of the subproblem for surgeon $s$. 
\end{property}
\begin{proof}
Lazy constraints $O-LC$ and $A-LC$ define the relation between $y_{sb}$ and $x_{pb}$ and this relation is not affected by the dual solution of the RMP as it is only projected into the objective function of the subproblem. Another potential difference between two runs of the subproblem entails the branching decisions that concern variables $y_{sb}$ ($y_{sb} = 1/0$, explained further in Section~\ref{section_BaP_Br}). Nevertheless, this cannot make any relation between $y_{sb}$ and $x_{pb}$ invalid (in the worst case, some lazy constraint will have no effect). As the other parameters of the subproblem remain the same, a lazy constraint determined in one run of the subproblem remains valid for any other run of the subproblem.
\end{proof}

The LCR procedure maintains a list of all lazy constraints (see the bottom part in Figure~\ref{fig:column_generation}) that are generated for each surgeon at run time of the branch-and-price algorithm. When the algorithm aims to generate a new pattern for surgeon $s$, the remembered lazy constraints are added to the subproblem. This reduces the number of callbacks that need to be generated and reduces the time the algorithm spends on solving subproblems. This technique applies to both $O-LC$ and $A-LC$.

Another speedup technique concerns whether the branch-and-price would generate only one new suitable pattern or more per single iteration of the column generation procedure. The former approach generates at most one suitable pattern per column generation iteration, i.e., the pricing procedure visits the subproblem of individual surgeons in logical order and is stopped whenever a suitable pattern has been found. The latter approach solves the subproblem for all surgeons and adds \textit{multiple suitable patterns} for all surgeons for whom one exists. This latter technique aims to reduce the number of column generation iterations and improve the convergence of the algorithm. 

\subsection{\label{section_BaP_BB}Branching and pruning decisions}
To obtain an (optimal) integer solution, we apply a branch-and-bound search. This method solves optimisation problems by breaking them into smaller subproblems through branching decisions, with each node relying on an optimal solution of the LPR obtained via column generation. This LPR solution serves as a bounding function, allowing the algorithm to prune nodes. Both these items are discussed below in more detail.

\subsubsection{\label{section_BaP_Br}Branching}
A natural choice for branching is to partition the solution space based on the original variables $y_{sb}$. The reason is that $y_{sb}$ variables are naturally decided in the RMP, while the patient assignment $x_{pb}$ comes from the callback function; thus, this assignment is less connected with the RMP. Taking into consideration the following property, the branching considers only variable $y_{sb}$.
\begin{property}
    If the branching decides all variables $y_{sb}$, the branch-and-price produces an integer bilevel feasible solution.
\end{property}
\begin{proof}
If the branching decides all variables $y_{sb}$, then this leads to the assignment of exactly one pattern to each surgeon, and therefore, all $x_{pb}$ will be binary as well since each pattern is associated with a bilevel assignment of patients, rendering a complete branching scheme. No further branching is required on variables $x_{pb}$ or to identify a bilevel feasible solution.
\end{proof}

Hence, whenever the column generation finds an optimal solution to the RMP that is fractional, the algorithm branches and 0-1 branching decisions are developed. Because $y_{sb}$ is not used in the RMP, its value is computed as $y_{sb} = \sum_{k \in \bar K_s} a_{kb}^s \theta_k^s$. The algorithm selects the variable whose value is closest to 0.5, i.e., the most fractional variable. The branching rule creates two nodes having smaller solution spaces, one where $y_{sb} = 1$ and the other where $y_{sb} = 0$. Branching decisions made in this and preceding nodes are stored in a branching history list $\mathit{BH}$ that is unique for every node in the branching tree. This list keeps track of the relevant branching decisions as triples $(s, b, e)$, where $e$ equals 1 or 0, distinguishing between the branch for $y_{sb} = 1$ and $y_{sb} = 0$, respectively.\\
Branching decisions impact both the subproblem and RMP. Branching constraints are imposed in the subproblem, such that new columns satisfy these conditions. Moreover, since each branching decision may conflict with one or more columns already included in the RMP, any columns that contradict the new branching decision are excluded from the RMP. As a result of this column management, the algorithm generates a default pattern that assigns blocks conforming to the triples $(s, b, 1) \in \mathit{BH}$ if there is no column for an arbitrary surgeon $s$. In this way, the algorithm guarantees that the RMP finds a feasible solution at all times.

\subsubsection{\label{section_BaP_Pr}Pruning}
The efficiency of branching depends on the effectiveness of solution space pruning, specifically through bounding. To prune nodes where the algorithm cannot find an optimal solution, we cut off nodes where the problem's linear relaxation has a worse or equal value of the objective function compared to the best solution found so far. To accelerate the finding of a good solution, the algorithm solves the so-called master heuristic in every node of the search tree \citep{Joncour2010}. Whenever the column generation finds the optimal value of the RMP, the master heuristic solves the RMP using MIP given the known set of columns $\bar K_s$ for every surgeon $s$ and binary domain conditions for the pattern variables $\theta_k^s \in \{0,1\}$, forcing the RMP to be solved to integrality. The yielded solution is compared to the best solution found so far, and if the new one is better, it is used as a new incumbent solution employed for the solution space pruning.

\subsection{\label{section_BaP_IH} Initial heuristic}

To prune non-promising branches early in the search and provide a strong initialisation of the RMP in the first iteration of the column generation at the root node, we derive a high-quality solution using a dedicated initial heuristic. This is a constructive heuristic that first assigns blocks to surgeons and then assigns patients such that the solution is bilevel feasible. The details of the heuristic, including the pseudocode and optimisation model, are presented in Online Appendix B.

The allocation of blocks establishes a block schedule, denoted by $\phi$, assigning blocks to surgeons for completing the surgery of patients, and defines the set of patients remaining to be scheduled for each surgeon $s$, represented by $\hat P_s$. Initially, $\phi$ is empty and $\hat P_s$ includes all patients in the set $P_s$ (\textit{Step 0}). After this initialisation, the heuristic adds in subsequent iterations surgeon-block assignments to the schedule. Each iteration is composed of several steps, explained in the following. In the first step, the heuristic assigns for every surgeon $s \in S$ and every possible block length $l \in L$ patients from $\hat P_s$ such that the leader objective is maximised (\textit{Step 1}). To that purpose, we solve for every surgeon and block length a knapsack problem considering the set of patients $\hat P_s$. Then, for every block $b$ with $\Gamma_b = l$ and $(s,b) \notin \phi$, we add $(s,b)$ to the set of candidate block assignments $\Omega$. Each assignment is associated with three parameters: the set of patients $\bar P_{sb}$ assigned to block $b$, and the contributions by the block to the leader objective $\bar \Delta_{sb}$ and $\bar \rho_{sb}$. The former represents the utilised capacity associated with block $b$ for surgeon $s$, i.e., $\bar \Delta_{sb} = \sum_{p \in \bar P_{sb}} \delta_p$, whereas the latter equals the priority of patients included in block $b$ for surgeon $s$, i.e., $\bar \rho_{sb} = \sum_{p \in \bar P_{sb}} \pi_{p}^{LP}$.
Next, we solve a block allocation model that aligns with the leader problem and assigns candidate block assignments $(s,b) \in \Omega$ to block schedule $\phi$, ensuring that at most one block is added per surgeon (\textit{Step 2}). The objective conforms to the objective of the leader problem but also considers an additional term to schedule the blocks as early as possible, such that blocks are ordered in a chronological manner and flexibility for allocating (large) blocks in later iterations of the heuristic is maximised. After this allocation step, the block schedule $\phi$ and set of patients $\hat P_s$ of patients remaining to be scheduled are updated (\textit{Step 3}). Steps 1 to 3 are repeated as long as the allocation model finds a feasible solution, i.e., $\phi$ is extended by a new assignment. These steps create a feasible schedule from a leader's point of view. \\
To generate a bilevel feasible solution, we discard the assignment of patients but keep the block schedule $\phi$. 
We solve the follower problem (\ref{eq:obj_follower_callback})-(\ref{eq:dom_follower_b}) for each surgeon given his/her assigned blocks (\textit{Step 4}). In the last step, we evaluate the constructed surgeon and patient schedule based on the leader objective (\ref{eq:obj_leader}) (\textit{Step 5}). Note that the assignment of patients obtained in \textit{Step 4} can be used in the LCR technique as each assignment is, in fact, a lazy constraint that can be added to the list of remembered lazy constraints.

\section{\label{section_results} Computational experiments}
In this section, we validate the computational performance of the proposed solution methodology and the added value of formulating the surgeon scheduling problem as a bilevel optimisation problem. In Section \ref{subsection_TestDesign}, we describe the relevant parameter settings and the characteristics of the test data set. Section
\ref{subsection_Results} benchmarks the proposed branch-and-price algorithm and compares it with the compact MIP model. In addition, we analyse the impact of implemented speed-up mechanisms and formulation of lazy constraints. Section \ref{subsection_gametheory} demonstrates the benefits of finding an equilibrium solution for different scenarios via calculation of the price of stability and price of decentralisation.

The imposed time limit for all experiments and solution methods is set at 1200 seconds per instance. The code is implemented in Python, and the tests are carried out on a 12th Gen Intel(R) Core(TM) i7-1255U processor @ 1.70 GHz and with 16GB of memory. The solver used to conduct the experiments is Gurobi 11.0. The source code of the described algorithms, problem instances, and their solutions are publicly available in the GitHub repository at \url{https://github.com/CTU-IIG/bilevel-branch-price-ors}.

\subsection{\label{subsection_TestDesign}Test design}
To validate the proposed solution methodology and problem definition, we consider the CHOIR data set of synthetic surgery scheduling instances developed by \cite{Leeftink2018}. The problem instances are constructed in an artificial manner, varying different settings and complexity indicators in a systematic and controlled way by means of an instance generator. In the following, we discuss the parametrisation of the employed instances in more details.

The instances are characterised by a planning horizon of 5 days ($|D|$) comprising 1, 2, or 4 operating rooms ($|R|$) and a load factor ($\mathrm{lf}$) of 1.5, 2.0, or 2.5 to induce necessary decisions in patient selection, making the trade-off with capacity utilisation. The load factor $\mathrm{lf}$ gives indication of the total expected surgery duration summed over all patients versus the available OR capacity. 
The daily capacity per room is 480 minutes, which equals 8 hours or 32 time slots, assuming time slots have a unit length of 15 minutes. Accordingly, the total capacity $C$ equals 32 $\times |R| \times |D|$. The lengths of operational OR blocks ($\Gamma_b$) equals multiples of two hours, ranging in $\{2, 4, 6, 8\}$. The possible start times of blocks are given by set $T = \{0, 2, 4, 6\}$ (in hours). Assuming the 15-minutes time slots as a time unit of the schedule, the set of possible block start times is $\{0, 8, 16, 24\}$ and the possible block lengths equal $\{8, 16, 24, 32\}$. The set of blocks $B_{d}$ comprises all blocks on day $d$ with a different start time and duration $\Gamma_b$, defining the set of overlapping blocks $O_{dt}$.
The set of surgeons $S$ involves a single surgeon group related to a particular discipline. The number of individual surgeons is a parameter that is varied and set to 10, 12, 14, or 20. We assume that the available rooms are dedicated to the involved surgeon group and individual surgeons are always available such that the set $N = \emptyset$. Surgeons can be assigned to at most $|D|$ blocks in the planning horizon ($v^h$) and to at most one block per day ($v^d$).

Surgeries are generated using the instance generator of~\cite{Leeftink2018}, which parametrises surgeries based on the patient case mix, desired $\mathrm{lf}$, and the number of OR days ($|R| \times |D|$). For our experiments, we have selected the case mix data set `1 RealLifeSurgeryTypesDatabase', which is based on real-life data from multiple academic and non-academic hospitals. The maximum deviation from the load factor is set to 0.025 per instance, and instance proximity $\varepsilon = 0.01$. The generator produces instances containing surgeries, such that each of them is described by a three-parameter lognormal distribution.
To preserve the load of each instance, the surgery duration $\delta_p$ is determined as an expected value of the distribution. The length of surgeries is divided by 15 minutes, i.e., the length of one time slot, and rounded to the nearest integer such that surgery duration can be measured in terms of time slots. Table \ref{tab:inst_char} provides an overview of the generated instances. Each row in the table describes a single data set characterised by $|S|$, $|R|$, and $\mathrm{lf}$. For each data set, we show the number of instances and summary statistics related to the number of surgeries and their duration.
Surgical cases are assigned to surgeons based on a uniform random basis, defining the set $P_s$ ($\forall s \in S$). The patient priorities of the surgeon head ($\pi_{p}^{LP}$) and individual surgeons ($\pi_p^{FP}$) are determined based upon a discrete uniform distribution considering the interval $[1,4]$, representing four different priority levels to differentiate between patients.

\begin{table}[!ht]
 \centering
 \small
\begin{tabular}{l | c | c c c | c c c}
					&  \textbf{\#instances} & \multicolumn{3}{c|}{\textbf{\#surgeries [-]}}	&	\multicolumn{3}{c}{\textbf{duration [time slots]}} \\
					&	& min & avg & max	&	min & avg & max \\
\hline
$|S|=20, |R|=2, \mathrm{lf}=2.0$	& 20 &		 81	&	93.9	&	103	&	1	&	6.8	&	30 	\\
$|S|=20, |R|=4, \mathrm{lf}=2.0$	& 20 &		 165	&	188.8	&	200	&	1	&	6.8	&	30 	\\
$|S|=12, |R|=1, \mathrm{lf}=1.5$	& 20 &		 31	&	36.3	&	45	&	1	&	6.7	&	30 	\\
$|S|=12, |R|=1, \mathrm{lf}=2.0$	& 20 &		 43	&	49.80	&	62	&	1	&	6.46	&	29 	\\
$|S|=12, |R|=1, \mathrm{lf}=2.5$	& 20 &		 55	&	61.6	&	69	&	1	&	6.5	&	30 	\\
$|S|=10, |R|=1, \mathrm{lf}=2.0$	& 20 &		 43	&	49.9	&	62	&	1	&	6.5	&	29 	\\
$|S|=14, |R|=1, \mathrm{lf}=2.0$	& 20 &		 43	&	49.9	&	62	&	1	&	6.5	&	29 	\\
\hdashline
$|S|=12, |R|=1, \mathrm{lf}=2.0$	& 30 &		 39	&	49.1	&	62	&	1	&	6.6	&	30	\\
\end{tabular}%
\caption{Instance data set characteristics per test set}
\label{tab:inst_char}%
\end{table}%

The objective function weights $\alpha$ and $\beta$ of the leader's objective equal to 1. In Section \ref{subsection_gametheory}, we investigate the impact of using different weight settings on the computational performance and the value of attaining an equilibrium solution. 
As the experiments with the MIP formulation are time-demanding, it is not possible to consider large sets of instances. Therefore, we consider 20 instances for each combination of $|S|$, $|R|$, and $\mathrm{lf}$ (Section \ref{subsection_benchmark}). For other experiments, not concerning the MIP, we consider the data set ($|S|=12, |R|=1,\mathrm{lf}=2.0$) with 30 instances to increase the accuracy of the experiments (sections \ref{subsection_algorithm} and \ref{subsection_gametheory}).
\color{black}

\subsection{\label{subsection_Results}Computational results}

\subsubsection{\label{subsection_benchmark}General performance and benchmark} 

In this section, we benchmark the general performance of the proposed branch-and-price algorithm ($BnP$) to solving the compact model (\ref{eq:obj_leader})-(\ref{eq:dom_follower}) using mathematical programming via the commercial solver Gurobi ($MIP$). Comparison is made accounting for different formulations of the lazy constraints that guide the leader problem towards a bilevel feasible solution (cf. Section \ref{subsec:LC}). Hence, devised algorithms include lazy constraints of type $O-LC$ or $A-LC$. Table~\ref{Table_Benchmark} displays the results related to the average performance of the different solution methods for the different instance sets indicated in Table~\ref{tab:inst_char}. The comparison is based on both the solution quality and the computational performance. The solution quality is measured via the yielded value of the leader objective ($F$), the objective value of the linear programming relaxation (LPR) in the root node ($F^{LPR}$), the relative optimality gap in the root node ($\%Gap = \frac{F-F^{LPR}}{F^{LPR}}$), the percentage of instances solved to optimality ($\%Opt$), and the percentage of instances for which a bilevel feasible solution has been found ($\%Feas$). The computational performance is reflected by the run time in seconds ($Time^T$). We highlight the best performance values in bold for each instance set and metric in the table.

\begin{sidewaystable}[htbp]
 \footnotesize
\centering
\begin{tabular}{l|cccccc|cccccc}
   & \multicolumn{6}{c|}{\textbf{BnP}} & \multicolumn{6}{c}{\textbf{MIP}}\\ 
   & \multirow{2}{*}{$F$} & \multirow{2}{*}{$F^{LPR}$} & \multirow{2}{*}{$\%Gap$} &\multirow{2}{*}{$\%Opt$} & \multirow{2}{*}{$\%Feas$}& \multirow{2}{*}{$Time^T$} & \multirow{2}{*}{$F$} & \multirow{2}{*}{$F^{LPR}$} & \multirow{2}{*}{$\%Gap$} &\multirow{2}{*}{$\%Opt$} & \multirow{2}{*}{$\%Feas$} & \multirow{2}{*}{$Time^T$} \\ 
 &   & &  & & & \\     \hline
\textbf{O-LC} &  & & && & \\
    \textrm{\quad  $|S|=10, |R|=1,\mathrm{lf}=2.0$ } & \textbf{45.3} & \textbf{45.1} & \textbf{0.4} & \textbf{90.0} & 100.0 & \textbf{155.2} & 47.3 & 26.9 & 78.9 & 0.0 & 100.0 & 1200.0 \\
    \textrm{\quad  $|S|=12, |R|=1,\mathrm{lf}=2.0$ } & \textbf{48.1} & \textbf{47.8} & \textbf{0.62} & \textbf{96.7} & 100.0 & 62.6 & 49.8 & 26.5 & 92.9 & 0.0 & 100.0 & 1200.0 \\
    \textrm{\quad  $|S|=14, |R|=1,\mathrm{lf}=2.0$ } & \textbf{51.3} & \textbf{51.1} & \textbf{0.4} & \textbf{95.0} & 100.0 & \textbf{72.5} & 52.6 & 26.9 & 98.6 & 0.0 & 100.0 & 1200.0 \\
    \textrm{\quad  $|S|=12, |R|=1,\mathrm{lf}=1.5$ } & \textbf{35.6} & \textbf{35.5} & \textbf{0.4} & \textbf{100.0} & 100.0 & 15.0 & 36.0 & 11.2 & 228.2 & 0.0 & 100.0 & 1200.0 \\
    \textrm{\quad  $|S|=12, |R|=1,\mathrm{lf}=2.5$ } & 68.3 & 64.7 & 0.2 & 90.0 & 100.0 & 154.6 & 70.6 & 46.9 & 50.5 & 0.0 & 100.0 & 1200.0 \\
    \textrm{\quad  $|S|=20, |R|=2,\mathrm{lf}=2.0$ } & \textbf{88.5} & \textbf{88.3} & \textbf{0.2} & \textbf{100.0} & 100.0 & \textbf{46.7} & 107.6 & 51.8 & 108.7 & 0.0 & 100.0 & 1200.0 \\
    \textrm{\quad  $|S|=20, |R|=4,\mathrm{lf}=2.0$ } & 136.1 & 102.8 & 0.3 & 50.0 & 100.0 & \textbf{841.0} & 195.8 & 95.6 & 109.5 & 0.0 & 20.0 & 1200.0 \\
    \hline
\textbf{A-LC} &  & & &&  &\\
    \textrm{\quad  $|S|=10, |R|=1, \mathrm{lf}=2.0$ } & \textbf{45.3} & \textbf{45.1} & \textbf{0.4} & \textbf{90.0} & 100.0 & 155.9 & 45.4 & 26.9 & 71.7 & 5.0 & 100.0 & 1148.5 \\
    \textrm{\quad  $|S|=12, |R|=1, \mathrm{lf}=2.0$ } & \textbf{48.1} & \textbf{47.8} & \textbf{0.6} & \textbf{96.7} & 100.0 & \textbf{55.6} & 48.2 & 26.5 & 86.5 & 3.3 & 100.0 & 1182.0 \\
    \textrm{\quad  $|S|=14, |R|=1, \mathrm{lf}=2.0$ } & \textbf{51.6} & \textbf{51.1} & \textbf{0.4} & \textbf{95.0} & 100.0 & 72.5 & 51.7 & 26.9 & 95.3 & 0.0 & 100.0 & 1200.0 \\
    \textrm{\quad  $|S|=12, |R|=1, \mathrm{lf}=1.5$ } & \textbf{35.6} & \textbf{35.5} & \textbf{0.4} & \textbf{100.0} & 100.0 & \textbf{12.2} & 35.7 & 11.2 & 225.8 & 10.0 & 100.0 & 1121.7 \\
    \textrm{\quad  $|S|=12, |R|=1, \mathrm{lf}=2.5$ } & \textbf{68.0} & \textbf{67.9} & \textbf{0.2} & \textbf{95.0} & 100.0 & \textbf{120.0} & 68.5 & 46.9 & 46.3 & 0.0 & 100.0 & 1200.0 \\
    \textrm{\quad  $|S|=20, |R|=2, \mathrm{lf}=2.0$ } & \textbf{88.5} & \textbf{88.3} & \textbf{0.2} & \textbf{100.0} & 100.0 & 52.2 & 95.3 & 51.8 & 84.7 & 0.0 & 100.0 & 1200.0 \\
    \textrm{\quad  $|S|=20, |R|=4, \mathrm{lf}=2.0$ } & \textbf{133.2} & \textbf{116.5} & 0.3 & \textbf{55.0} & 100.0 & 874.1 & 168.9 & 95.6 & 76.7 & 0.0 & 100.0 & 1200.0 \\
    \hline
\end{tabular}
 \caption{Benchmark comparison between alternative solution approaches and lazy constraint formulations}
 \label{Table_Benchmark}
\end{sidewaystable}
 \normalsize

Table~\ref{Table_Benchmark} reveals that the proposed branch-and-price is superior to MIP in terms of both the solution quality of yielded solutions and computational performance, independent of the type of lazy constraint implemented. Results demonstrate that the branch-and-price consistently attains solutions of higher quality ($F$), a large percentage of solutions for which optimality is proven within the time limit ($\%Opt$), and substantially smaller run times ($Time^T$). This is predominantly because of the tight LPR in the root node realised via the branch-and-price method and implemented decomposition. Main difference is that the generated columns in the branch-and-price method represent bilevel feasible solutions from the perspective of individual surgeons. This cannot be assured by the MIP, which explains the large optimality gaps in the root node and, correspondingly, the large run times and low percentage of instances for which optimal solutions have been proven. 

When comparing the type of lazy constraints included, we observe that only for a few of the instances, the MIP solver can prove the optimality of a solution when lazy constraints of type \textit{A-LC} are implemented. No such proof of optimality is achieved when using lazy constraints of type \textit{O-LC}. Moreover, the quality $F$ of yielded bilevel solutions is consistently better for lazy constraints of type \textit{A-LC}. This can be explained as \textit{A-LC}s outperform \textit{O-LC}s in reducing symmetries because the former focus on fixing patient assignments, while the latter operate primarily based on the objective function and leave some room for alternative solutions satisfying the objective requirement of the follower, especially when the instances consider multiple number of rooms. This is observed for instance sets with $|R|$ = 2 or 4. On the other hand, the performance of the branch-and-price method is only slightly impacted by the type of lazy constraint implemented because of the tight bound provided by the LPR. This is foremost visible in the run times of the algorithm, whereas the difference in the objective value $F$ and percentage of instances solved to optimality is limited. For the branch-and-price, lazy constraints of type \textit{A-LC} are more effective, consistently producing tighter lower bounds in the root node ($F^{LPR}$). This is because for some instances in instance sets ($|S|=12, |R|=1, \mathrm{lf}=2.5$) and ($|S|=20, |R|=4, \mathrm{lf}=2$) the column generation procedure at the root node fails to converge to an optimal LPR solution within the time limit when lazy constraints of type \textit{O-LC} are applied. Correspondingly, run times are in general smaller when the branch-and-price includes lazy constraints of type \textit{A-LC} for instance sets featuring only a single room. However, for instances considering multiple rooms, the branch-and-price records smaller run times when lazy constraints of type \textit{O-LC} are implemented. Detailed results show that the latter is attributed to the smaller number of nodes that need to be explored, despite lazy constraints of type \textit{A-LC} induce less column generation iterations per node and require shorter times for solving subproblems.

\subsubsection{\label{subsection_algorithm}Impact of algorithm improvement mechanisms}

In order to assess the impact of the introduced acceleration mechanisms, we compare different versions of the proposed algorithm. Starting from the best-algorithm configuration, each version leaves out one specific mechanism keeping the rest of the algorithm unchanged. In this regard, we exclude either the initial heuristic (\textit{w/o InH}), the generation of multiple patterns per pricing iteration and stop the pricing step whenever a single promising pattern with positive reduced cost has been found (\textit{w/o MuP}), or the remembering of lazy constraints (\textit{w/o LCR}). In addition, we further display the detailed computational implications of generating lazy constraints of types \textit{A-LC} and \textit{O-LC} to cut off bilevel infeasible solutions. Table~\ref{tab:Speed-up} presents the impact of these settings on the computational performance of the proposed algorithm, averaging the results over the instance set featured by 12 surgeons, 1 OR, and a load factor equal to 2. We report in this table information related to (i) the functioning of the algorithm, i.e., the number of column generation iterations ($\#CGI$), columns added to the master problem ($\#Cols$), lazy constraints generated ($\#LCs$), callbacks ($\#CBs$), and nodes explored in the branch-and-bound tree ($\#Nodes$); (ii) the obtained solution quality, i.e., the relative optimality gap ($\%Gap$), the percentage of instances for which a bilevel feasible solution has been found ($\%Feas$), and the percentage of instances solved to optimality ($\%Opt$); and (iii) the computational performance, i.e., the run time spent in the callback function to derive lazy constraints ($Time^{\mathit{CB}}$), the run time to solve the complete pricing step (including the callback time) ($Time^{\mathit{SP}}$), the run time to solve the master problem ($Time^{\mathit{MP}}$), and the total required run time ($Time^{\mathit{T}}$).

\begin{table}[htbp]
 \centering
 \small
 \begin{tabular}{l| c c c c | c}
 & \multicolumn{4}{c|}{\textit{A-LC}} & \multicolumn{1}{c}{\textit{O-LC}}\\
 
\multicolumn{1}{r|}{} & \multicolumn{1}{c}{\textit{w/o}} & \multicolumn{1}{c}{\textit{w/o}} & \multicolumn{1}{c}{\textit{w/o}} &\\

\textbf{} & \multicolumn{1}{c}{\textit{InH}}& \multicolumn{1}{c}{\textit{MuP}} & \multicolumn{1}{c}{\textit{LCR}} & \multicolumn{1}{c|}{\textit{All}}&  \multicolumn{1}{c}{\textit{All}}\\
 \hline 
\textbf{Algorithm performance} & &  &  &  &  \\
\hspace{3mm}$\#CGI$ & 69.8 & 317.0 & 67.0 & 68.0 & \textbf{65.9} \\
\hspace{3mm}$\#Cols$ & 291.7 & 305.7 & 256.4 & 257.4 & \textbf{250.9} \\
\hspace{3mm}$\#LCs$ & 41.4 & 42.6 & 395.6 & \textbf{36.5} & 37.1 \\
\hspace{3mm}$\#CBs$ & 2460.2 & 3459.2 & 2666.3 & 2290.8 & \textbf{2158.6} \\
\hspace{3mm}$\#Nodes$ & 18.3 & 12.1 & 18.5 & 18.2 & 20.6 \\
\hdashline 
\textbf{Solution quality} & &  &  &  & \\
\hspace{3mm}$\%Gap$ & 0.6 & 0.6 & 0.6 & 0.6 & 0.6 \\
\hspace{3mm}$\%Feas$ & 100.0 & 100.0 & 100.0 & 100.0 & 100.0 \\
\hspace{3mm}$\%Opt$ & 96.7 & 96.7 & 96.7 & 96.7 & 96.7 \\
\hdashline
\textbf{Run time} & &  &  &  & \\
\hspace{3mm}$Time^{\mathit{CB}}$ & 3.6 & 4.6 & 6.0 & 3.0 & \textbf{2.3} \\
\hspace{3mm}$Time^{\mathit{SP}}$ & 24.5 & 66.7 & 26.9 & \textbf{24.3} & 30.9 \\
\hspace{3mm}$Time^{\mathit{MP}}$ & 32.2 & 70.5 & \textbf{30.1} & 31.2 & 31.7 \\
\hspace{3mm}$Time^T$ & 56.7 & 137.3 & 57.0 & \textbf{55.6} & 62.6 \\
  \end{tabular}%
   \caption{Impact of the introduced speed-up mechanisms and lazy constraints for instance set ($|S|=12, |R|=1, \mathrm{lf}=2.0$)}
\label{tab:Speed-up}%
\end{table}%

Table~\ref{tab:Speed-up} reveals that the implemented speed-up techniques enhance the performance of the branch-and-price algorithm across various metrics, whereas there is no difference in obtained solution quality for the considered instance set. All acceleration mechanisms improve the run time of the proposed algorithm ($Time^T$).
Omitting the generation of multiple patterns (\textit{w/o MuP}) is most impactful, increasing the run times from 55.6 to 137.3 seconds. This is explained by the substantially larger number of iterations required by column generation ($\#CGI$) and its slower convergence to an optimal LPR solution as approximately 20\% more columns ($\#Cols$) with suitable reduced cost need to be generated (305.7 versus 257.4) to prove optimality of the LPR across various nodes. Also, because a substantially larger number of pricing problems is visited, which is a function of the number of column generation iterations, the callback function ($\#CBs$) is invoked many times more (3459.2 versus 2290.8). Incorporating the mechanism to remember lazy constraints has only a modest impact on the overall run time, as it primarily affects the solving of the pricing problem, which can be handled efficiently for the problem under study. The mechanism accelerates the solving of the pricing problem, i.e., $Time^{SP}$ reduces from 26.9 (\textit{w/o LCR}) to 24.3 (\textit{All}) seconds. This stems from the substantially smaller amount of time that needs to be spent on solving callback functions ($Time^{CB}$) to check bilevel feasibility, encompassing a 50\% reduction from 6.0 to 3.0 seconds. The number of relevant lazy constraints ($\#LCs$) reduce by more than 90\%, from 395.6 to 36.5, and the number of callbacks ($\#CBs$) decreases by approximately 10\% from 2666.3 to 2290.8. 
The initial heuristic has also only little impact on the computational performance, reducing slightly the number of columns with suitable reduced cost generated via column generation ($\#Cols$) and the number of callbacks ($\#CBs$), leading overall to a small improvement in run time from 56.7 to 55.6 seconds ($Time^T$) when the initial heuristic is included. However, the main advantage of the initial heuristic is that it guarantees the identification of a bilevel feasible solution, even in cases where the column generation algorithm struggles to converge (see, e.g., instance sets ($|S|=12, |R|=1, \mathrm{lf}=2.5$) and ($|S|=20, |R|=4, \mathrm{lf}=2$) with the implementation of \textit{O-LC} in Table~\ref{Table_Benchmark}).\\ 
A detailed analysis of the algorithm's performance when including lazy constraints either of type \textit{A-LC} or \textit{O-LC} reveals that the difference lies foremost in the time for solving subproblems, whereas the number of pricing problems visited is larger when lazy constraints of type \textit{A-LC} are implemented. Incorporating \textit{A-LC}s reduces the required run time $Time^{SP}$ by approximately 20\%, with an average run time for solving the pricing problem decreasing from 30.9 (\textit{O-LC}) to 24.3 (\textit{A-LC}) seconds, although the number of incorporated lazy constraints ($\#LCs$) of both types is approximately the same. This is explained as constraints of type \textit{A-LC} ensure a tighter formulation of the subproblem, which can be solved in smaller run times. 

\subsection{\label{subsection_gametheory}Value of game-theoretic approach}
In this section, we assess the value of attaining an equilibrium solution and the consequences when surgery schedules are composed in a decentralised, independent manner by the individual surgeons. For that purpose, we quantify the impact on the leader's objective via comparison between solution values of an optimal equilibrium (or bilevel) solution and an optimal (de-)centralised solution via following metrics:

\begin{itemize}
\item	The \textit{Price of Stability} ($PoS$) measures how the efficiency of a system degrades due to the required equilibrium associated with an optimal bilevel solution \citep{Paccagnan2022}. $PoS$ is formalised as

\footnotesize
\begin{equation} \label{eq:LPoS}
PoS = \frac{min_{e' \in E'} F(e')}{min_{e \in E} F(e)}\end{equation}
\normalsize

with $E$ the set of all solutions ($E = \{0,1,\ldots, |E|-1\}$, with index $e$) and $E'$ the set of equilibrium solutions ($E' = \{0,1,\ldots, |E'|-1\}$, with index $e'$). Equation (\ref{eq:LPoS}) defines $PoS$ as the ratio between the optimal solution values for the bilevel problem under study and the so-called centralised problem, i.e.,  \vspace{-2mm}
\begin{itemize}[noitemsep]
\item [-] The \textit{bilevel problem} corresponds to model (\ref{eq:obj_leader})-(\ref{eq:dom_follower}), which is solved using the described branch-and-price algorithm. A(n) (optimal) bilevel solution yields an equilibrium as none of the surgeons can improve their own objective without having to change the allocation of the OR blocks, which is determined by the leader.
\item [-] The \textit{centralised problem} is represented by model (\ref{eq:obj_leader})-(\ref{eq:dom_follower}), excluding the objective of the surgeons (eq. (\ref{eq:obj_follower})). This model determines an optimal centralised solution, which provides the best possible objective value for the leader while disregarding the incentives of the individual surgeons.
The formulation of the centralised problem is presented in Online Appendix C.
\end{itemize}

\item The \textit{Price of Decentralisation} ($PoD$) measures how the efficiency of a system may degrade due to selfish behaviour of its agents from the perspective of the leader. $PoD$ is formalised as

\footnotesize
\begin{equation} \label{eq:LPoA}
PoD = \frac{F(e\in E|e= \argmax_e{\sum_{s\in S}f_s(e)})}{min_{e \in E} F(e)}.\end{equation}
\normalsize

Equation (\ref{eq:LPoA}) defines $PoD$ as the ratio between the optimal solution values for the so-called decentralised problem and the centralised problem. The \textit{decentralised problem} is represented by model (\ref{eq:obj_leader})-(\ref{eq:dom_follower}), with the objective modified to maximise the sum of the followers' objectives (i.e., $\sum_{s \in S} f_s$ or $\sum_{s \in S}\sum_{p\in P_s} \pi_p^{FP} \sum_{b\in B} x_{pb}$) replacing eq. (\ref{eq:obj_leader}), and omitting constraint  (\ref{eq:obj_follower}). This model determines an optimal decentralised solution, which inherently leads to an equilibrium as it maximises the reward of the individual surgeons. Nevertheless, the solution in (\ref{eq:LPoA}) is evaluated by the leader objective $F$. The formulation of the decentralised problem is presented in Online Appendix C.

$PoD$ indicates how bad a solution may turn out if the leader allows surgeons to plan the surgeries while the surgeons disregard the leader's objective. Note that $PoD$ has been selected over the Price of Anarchy \citep{Paccagnan2022}, as the latter makes comparison to the worst equilibrium solution, which corresponds to an empty schedule and is irrational.




\end{itemize}

These prices are calculated for different scenarios in order to value attaining a game-theoretic equilibrium in different settings, depending upon (i) the patient priorities set by the surgeon head versus those of the individual surgeons, and (ii) the objective function structure of the leader determined by weights $\alpha$ and $\beta$. We consider the following scenarios:

\textit{Scenarios depending upon patient priorities ($\pi_p^{FP}$, $\pi_{p}^{LP}$)}
\begin{itemize}[noitemsep]
\item [-] \textbf{Scenario 1} ($\pi_p^{FP} = \pi_{p}^{LP} = 1$): Priorities are equal to 1 for all patients, maximising the number of patients scheduled. The department head sets patient priorities in accordance to the individual surgeons.
\item	[-] \textbf{Scenario 2} ($\pi_p^{FP} = \pi_{p}^{LP} \in  [1, 4]$ (random)): Priorities can vary between patients. The department head aligns these priorities with those established by the individual surgeons. 
\item	[-] \textbf{Scenario 3} ($\pi_p^{FP} = 1$; $\pi_{p}^{LP} \in  [1, 4]$ (random)): Priorities set by the department head can vary between patients. These patient priorities are different than the ones set by the individual surgeons, who set all patient priorities equal.
\item	[-] \textbf{Scenario 4} ($\pi_p^{FP} \in  [1, 4]$ (random); $\pi_{p}^{LP} = 1$): Priorities set by individual surgeons can vary between patients, whereas the surgeon head sets all patient priorities equal, maximising the number of patients treated.
\item [-] \textbf{Scenario 5} ($\pi_p^{FP} \in  [1, 4]$ (random); $\pi_{p}^{LP} \in  [1, 4]$ (random)): Priorities can vary between patients for both the department head and individual surgeons.  
\end{itemize}

\textit{Scenarios depending upon objective weights ($\alpha$, $\beta$)}\\
To evaluate the impact of the weights relative to the leader objective, i.e., minimising the idle OR time ($\alpha$) and the priority penalty associated with patients that are not included in the surgical case planning ($\beta$), we experiment with different absolute weight settings, i.e., ($\alpha$, $\beta$) = (1,0), (0,1), and (1,1), considering only the first objective, second objective, or both objectives with equal weights, respectively.

Table \ref{Table_gamevalue} presents the results related to $PoS$ and $PoD$ for the different scenarios tested. As the weight setting ($\alpha$, $\beta$) = (1,0) does not account for the patient priority in the leader's objective, we consider only Scenarios 1 and 5, having all priorities set by the individual surgeons equal or distinct. Accordingly, we explore in total $5\times 2 + 2 = 12$ scenarios. Without loss of generality, we consider in this experiment only instance set ($|S|=12, |R|=1, \mathrm{lf}=2.0$) such that only optimal results are reported, giving unbiased insight into the efficiency of an equilibrium solution, which cannot be guaranteed for larger instances due to the imposed time limit as stop condition (note: experiments have been conducted without the initial heuristic to ensure consistency between the bilevel and (de-)centralised optimisation runs). The table displays average results for both the computational performance and solution quality relative to the best equilibrium solution and best (de-)centralised solution. The solution quality is measured via the leader's objective ($F$, see eq. (\ref{eq:obj_leader})), the leader's priority of planned patients ($P^L = \sum_p \sum_b \pi_{p}^{LP} x_{pb}$), the OR utilisation ($U = \frac{\sum_p \sum_b \delta_p x_{pb}}{C}$), and the sum of the followers' objectives, i.e., the followers' priority of scheduled patients ($\sum_s f_s$, see eq. (\ref{eq:obj_follower})). In addition, we report $PoS$ and $PoD$. The computational performance is reflected by the run time in seconds ($Time^T$). The main observation gained from the experiment is the following:

\begin{sidewaystable}[htbp]
 \footnotesize
\centering
\begin{tabular}{l|ccccc|ccccc|ccccc|cc}
\multirow{2}{*}{} &  \multicolumn{5}{c|}{{\textbf{Best equilibrium solution}}} & \multicolumn{5}{c|}{{\textbf{Best decentralised solution}}} & \multicolumn{5}{c|}{{\textbf{Best centralised solution}}} & \multirow{2}{*}{$PoS$} & \multirow{2}{*}{$PoD$} \\
& \multirow{1}{*}{$F$}   & \multirow{1}{*}{$P^L$} & \multirow{1}{*}{$U$} & \multirow{1}{*}{$\sum_s f_s$} & \multirow{1}{*}{$Time^T$} & \multirow{1}{*}{$F$}   & \multirow{1}{*}{$P^L$} & \multirow{1}{*}{$U$} & \multirow{1}{*}{$\sum_s f_s$} & \multirow{1}{*}{$Time^T$}& \multirow{1}{*}{$F$}   & \multirow{1}{*}{$P^L$} & \multirow{1}{*}{$U$} & \multirow{1}{*}{$\sum_s f_s$} & \multirow{1}{*}{$Time^T$} &&\\ \hline
 \textbf{($\alpha$, $\beta$) = (1,0)} &  & & && & &&&&&&&&& && \\
    \textrm{\quad  Scenario 1} & \textcolor{black}{0.500} & 25.3 & $\approx$1 & 25.3 & 17.4 & \textcolor{black}{25.067} & 31.8 & 0.84 & 31.8 & 6.5 & \textcolor{black}{0.067} & 24.1 & $\approx$1 & 24.1 & 12.7 & 7.500 & 376.000 \\
    \textrm{\quad  Scenario 5} & \textcolor{black}{1.367} & 64.0 & 0.99 & 71.7 & 65.8 & \textcolor{black}{21.533} & 82.5 & 0.87 & 96.2 & 8.9 & \textcolor{black}{0.067} & 58.9 & $\approx$1 & 63.8 & 12.2 & 20.500 & 323.000 \\
\hdashline
 \textbf{($\alpha$, $\beta$) = (0,1)} &  & & && & &&&&&&&&& && \\
    \textrm{\quad  Scenario 1} & \textcolor{black}{17.233} & 31.8 & 0.85 & 31.8 & 8.2 & \textcolor{black}{17.233} & 31.8 & 0.84 & 31.8 & 5.7 & \textcolor{black}{17.233} & 31.8 & 0.84 & 31.8 & 6.4 & 1.000 & 1.000 \\
    \textrm{\quad  Scenario 2} & \textcolor{black}{36.833} & 95.8 & 0.87 & 95.8 & 11.5 & \textcolor{black}{36.833} & 95.8 & 0.87 & 95.8 & 7.8 & \textcolor{black}{36.833} & 95.8 & 0.87 & 95.8 & 10.6 & 1.000 & 1.000 \\
    \textrm{\quad  Scenario 3} & \textcolor{black}{36.900} & 95.7 & 0.87 & 31.1 & 11.4 & \textcolor{black}{46.633} & 86.0 & 0.85 & 31.8 & 5.9 & \textcolor{black}{36.833} & 95.8 & 0.87 & 31.0 & 9.3 & 1.002 & 1.266 \\
    \textrm{\quad  Scenario 4} & \textcolor{black}{17.267} & 31.8 & 0.84 & 86.5 & 8.4 & \textcolor{black}{18.033} & 31.0 & 0.87 & 95.8 & 7.7 & \textcolor{black}{17.233} & 31.8 & 0.84 & 86.0 & 6.1 & 1.002 & 1.046 \\
    \textrm{\quad  Scenario 5} & \textcolor{black}{36.667} & 89.6 & 0.86 & 89.0 & 10.6 & \textcolor{black}{43.766} & 82.5 & 0.87 & 96.2 & 9.1 & \textcolor{black}{35.133} & 91.1 & 0.87 & 86.7 & 8.4 & 1.044 & 1.246 \\
\hdashline
 \textbf{($\alpha$, $\beta$) = (1,1)} &  & & && & &&&&&&&&& && \\
    \textrm{\quad  Scenario 1} & \textcolor{black}{21.367} & 29.7 & 0.99 & 29.7 & 18.6 & \textcolor{black}{42.233} & 31.8 & 0.84 & 31.8 & 5.8 & \textcolor{black}{21.233} & 29.4 & 0.99 & 29.4 & 16.8 & 1.006 & 1.989 \\
    \textrm{\quad  Scenario 2} & \textcolor{black}{47.633} & 91.6 & 0.96 & 91.6 & 59.4 & \textcolor{black}{57.967} & 95.8 & 0.87 & 95.8 & 7.8 & \textcolor{black}{47.067} & 91.6 & 0.96 & 91.6 & 35.6 & 1.012 & 1.232 \\
    \textrm{\quad  Scenario 3} & \textcolor{black}{47.333} & 91.3 & 0.96 & 30.5 & 63.5 & \textcolor{black}{71.133} & 86.0 & 0.85 & 31.8 & 5.9 & \textcolor{black}{47.067} & 91.6 & 0.96 & 30.3 & 35.4 & 1.006 & 1.511 \\
    \textrm{\quad  Scenario 4} & \textcolor{black}{23.000} & 29.2 & 0.98 & 83.2 & 102.3 & \textcolor{black}{38.700} & 31.0 & 0.87 & 95.8 & 7.3 & \textcolor{black}{21.233} & 29.4 & 0.99 & 79.4 & 18.5 & 1.083 & 1.823 \\
    \textrm{\quad  Scenario 5} & \textcolor{black}{48.067} & 85.1 & 0.97 & 87.5 & 58.6 & \textcolor{black}{65.300} & 82.5 & 0.87 & 96.2 & 7.6 & \textcolor{black}{45.167} & 86.5 & 0.97 & 83.9 & 100.9 & 1.064 & 1.446 \\
    \end{tabular}
 \caption{The efficiency of finding equilibrium solutions ($PoS$ and $PoD$) for instance set ($|S|=12, |R|=1, \mathrm{lf}=2.0$) }
 \label{Table_gamevalue}
\end{sidewaystable}
 \normalsize

\begin{observation}
$PoS$ and $PoD$ are inversely related to the correlation between the leader and follower objectives. The larger the correlation is, the closer $PoS$ and $PoD$ are to 1.
\end{observation}
\vspace{-4mm}
\begin{proof}[\unskip\nopunct]
Table \ref{Table_gamevalue} reveals that the worse the alignment is between the leader and follower objectives, the higher the values are for $PoS$ and $PoD$. This implies that in particular cases it is more relevant to consider a bilevel optimisation framework as selfish behaviour of individual surgeons degrades the solution quality considerably, but - at the same time -  this obtained stability comes at a higher cost from the perspective of the leader. 
\end{proof}

In the following, we provide insights into the values of these metrics for different objective weight settings and defined patient priorities.

\textit{Objective weight setting: $\alpha = 0$ and $\beta = 1$}\\
In this case, both the objectives of the leader and followers are related to the selection of individual patients only and there is no interference with the objective of minimising the underutilisation. Consequently, attained utilisation rates are lower, ranging between 84 and 88\%. An analysis of the results leads to following observations:

\begin{observation}
When patient priorities among leader and follower are identical and constitute the sole objective, $PoS$ and $PoD$ are equal to 1. 
\end{observation}
\vspace{-4mm}
\begin{proof}[\unskip\nopunct]
The results for scenarios 1 and 2, for which the leader and follower priorities are identical, show that inherently $PoS$ and $PoD$ equal 1, which imply that the best centralised, decentralised and bilevel feasible solutions are naturally identical to each other and there is no need to consider the concept of bilevel feasibility. 
\end{proof}

\begin{observation}
When patient priorities for the leader and followers are distinct, $PoS$ is higher than 1, and increases slightly when objectives are less correlated.
\end{observation}
\vspace{-4mm}
\begin{proof}[\unskip\nopunct]
In scenarios 3 and 4, where either the leader or the follower assumes equal priorities of 1 to maximise the number of planned surgeries, the objective values of the best centralised and bilevel feasible solutions are highly comparable, i.e., $PoS$ is slightly different from 1. This results from the large (positive) correlation between defined leader and follower priorities. However, when differing priorities are considered, as in scenario 5, a higher price is charged to achieve a stable solution as the correlation between the leader and follower objectives decreases. Specifically, $PoS$ increases to 1.044, indicating that the leader's objective $F$ in an optimal bilevel solution is worse than in a best centralised solution \textcolor{black}{(36.667 versus 35.133)}.
\end{proof}

\begin{observation}
When the leader considers varying priorities for selecting patients, different from the priorities of the followers, the best centralised and decentralised solutions are considerably different and $PoD$ is significantly higher than 1.
\end{observation}
\vspace{-4mm}
\begin{proof}[\unskip\nopunct]
In scenarios 3 and 5, the sets of surgeries selected by the leader and followers are different in the best centralised and decentralised solutions due to distinct patient priorities. This implies that the solution degrades significantly when individual surgeons independently create their own patient schedules ($PoD$ = 1.266 and 1.246). This is not prevalent for scenario 4, where the leader maximises the number of patients treated. In that case, $PoD$ is smaller, with a value of 1.046, as the head surgeon sets all priorities equal to 1 and remains indifferent to the specific patients chosen. As noted, $PoS$ is low for scenarios 3 and 4, indicating that a stable solution, which prevents selfish behaviour of surgeons, can be achieved with only a minor reduction in solution quality.
\end{proof}

\textit{Objective weight setting: $\alpha = 1$ and $\beta = 0$}\\
In this case, the leader objective considers only the OR underutilisation and no priorities for selecting patients, which leads to the following observation:

\begin{observation}
Limited alignment between leader and follower objectives leads to values for $PoS$ and $PoD$ that are substantially larger than 1. Achieving a stable solution involves a substantial cost; however, this is necessary to avoid significantly worse outcomes for the leader that could arise if individual surgeons were to independently determine their own schedules. Hence, a coordination mechanism to contain the followers' behaviour is indispensable.
\end{observation}
\vspace{-4mm}
\begin{proof}[\unskip\nopunct]
When the leader objective considers only the OR underutilisation, both $PoS$ and $PoD$ take substantially larger values. This occurs due to the limited alignment between the leader and follower objectives. The former relates only indirectly to the selection of (an aggregated set of) patients, resulting in reduced coordination. The $PoD$ attains for both scenarios 1 and 5 extremely large values, i.e., decentralised optimisation results in objective values $F$ equal to \textcolor{black}{25.067} and \textcolor{black}{21.533} with lower utilisation rates, ranging between 84\% and 87\%, whereas the optimal centralised solutions denotes an objective value of \textcolor{black}{0.067} and utilisation rates close to 100\%. As a result, the values for $PoD$ amount to 376 (=25.067/0.067) and 323 (=21.533/0.067) for scenarios 1 and 5, respectively. In contrast to the values for the leader objective, the attained solution quality for the follower objective is far smaller for optimised centralised versus decentralised solutions. Furthermore, $PoS$ records also larger values than for any other objective weight setting, which is equal to 7.5 when surgeons try to maximise the number of cases treated, considering patient priorities equal to 1, and 20.5 when patient priorities are distinct among patients.  
\end{proof}

\textit{Objective weight setting: $\alpha = 1$ and $\beta = 1$}\\
The case the leader objective considers both objectives leads to following observation:

\begin{observation}
Considering an additional objective that is only indirectly related to the follower objective (i.e., minimising the OR underutilisation), alongside of the objective involving the selection of specific patients, increases the values for $PoS$ and $PoD$. Values for $PoD$ are especially larger when the leader sets all priorities equal to 1. When both parties postulate the same priorities or followers maximise the number of patients treated, $PoS$ is close to 1 and only a small objective value needs to be sacrificed to realise a stable outcome.
\end{observation}
\vspace{-4mm}
\begin{proof}[\unskip\nopunct]
The values for $PoS$ and $PoD$ lie in between the prices observed for weight settings ($\alpha, \beta$) = (1,0) or (0,1). Results reveal that $PoS$ attains the largest values for scenarios having distinct patient priorities set by the individual surgeons, i.e., for scenarios 2 and especially 4 and 5. Scenarios for which the followers try to maximise the number of cases treated with $\pi_p^{FP}$ equal to 1 ($\forall p \in P$) denote considerably smaller values for $PoS$, close to 1, as individual surgeons are indifferent to the specific patients selected by the head surgeon. The largest values for $PoD$ are observed for scenarios 1 and 4, where the leader priorities for selecting patients all equal to 1 as the latter gives relatively more weight to the objective minimising the OR underutilisation. The $PoD$ is smallest when both the leader and followers align their priorities, varying across patients, i.e., scenario 2.
\end{proof}

\section{\label{section_conclusion}Conclusion and future study}
In this paper, we have studied a bilevel OR scheduling problem defining the surgeon schedule for the upcoming short-term horizon. This schedule results from a negotiation process between the surgeon head and individual surgeons. The former is tasked with distributing OR time across surgeons, whereas the latter plan surgical cases in an independent manner. Involved parties have their own objectives and the behaviour of the individual surgeons might degrade the schedule quality envisaged by the surgeon head. The presented model generates a surgeon schedule and advance planning of surgeries that is a stable solution, preventing schedule changes by the individual surgeons seeking to improve their own objectives. We have proposed an exact branch-and-price solution approach that searches the bilevel solution space in an efficient manner, relying on several problem properties. Most important one is the considered decomposition of the problem, which defines subproblems per surgeon and a master problem aligning the schedules of individual surgeons. This allows the generation of bilevel feasible patterns for individual surgeons and, consequently, the generation of tight bounds for the LPR of the problem under study. Computational experiments demonstrate the superiority of the presented approach and provide managerial insights. The main observation is that the weaker the correlation between the leader and follower objectives is, the poorer the outcomes are when individual surgeons independently compose their schedules. This highlights the relevance of addressing the composition of surgeon schedules via bilevel optimisation, despite the increased complexity and decrease in objective value required to align the surgeon head and individual surgeons with respect to the final patient schedule.

This research considers a novel, fundamental problem in OR planning and scheduling opening multiple future research avenues. The problem definition can be extended by considering additional well-known features inherent to the OR environment complicating the modelling of the surgeons' behaviour such as introducing uncertainty relative to the duration and/or priority of surgeries. In that case, the leader has no longer perfect knowledge about the surgeons' optimisation problem and stochastic and (distributionally) robust optimisation techniques become relevant. Other extensions may relate to the modelling of the behaviour between (groups of) surgeons, involving specific patient-surgeon relationships (e.g., multiple surgeons can be required/candidate to perform a patient's surgery) or the sharing of resources such as expensive equipment or OR time, which is relevant in an open booking or modified block booking framework. In the context of the latter, the problem may assume an upper-bounded overtime, defined for every OR and day, such that its use can be negotiated among the surgeons. Then, the assignment of patients depends not only on the time allocated by the leader but also on the willingness of other surgeons operating in the same OR day to use the overtime. This means that when one surgeon decides to leave the overtime to the others, (an)other surgeon(s) may want to modify the assignment of their patients since using the overtime may increase their objective function. This shows two research opportunities: (i) a need to extend the model of surgeons' behaviour, namely their pay-off function, and (ii) a need to revise the above-described decomposition such that the master problem can guarantee bilevel feasibility of the solution regardless whether overtime is used or not.

\section{Acknowledgements}

This work was supported by the Grant Agency of the Czech Republic under the Project GACR 22-31670S, and co-funded by the European Union under the project ROBOPROX - Robotics and Advanced Industrial Production (reg. no. CZ.02.01.01/00/22\_008/0004590). 

\bibliographystyle{apalike}
\bibliography{references}

\end{document}